\documentclass[preprint]{elsarticle}

\usepackage{graphicx}
\usepackage{latexsym}
\usepackage[abs]{overpic}
\usepackage{wrapfig}
\usepackage{amssymb}
\usepackage{amsmath}

\newcommand{\eq} {equation}

\newcommand{\eqa} {eqnarray}

\newcommand{\NN} {\mbox {$\nonumber$}}

\newcommand{\RA}  {\mbox {$\rightarrow$}}

\newcommand\non{\nonumber \\}

\def\mN{\mathcal{N}}
\def\Tr{\mathrm{Tr}}
\def\tr{\mathrm{tr}}
\def\tX{\tilde{X}}
\def\tsig{\tilde{\sigma}}
\def\sig{{\sigma}}

\def\R{{\mathbb{R}}}
\def\Z{{\mathbb{Z}}}
\def\P{{\mathbb{P}}}

\newcommand{\be}{\begin{equation}}
\newcommand{\ee}{\end{equation}}
\newcommand{\bea}{\begin{eqnarray}}
\newcommand{\eea}{\end{eqnarray}}
\newcommand{\bfJ} {\mathbf{J}}
\newcommand{\bbJ} {\mathbb{J}}

\journal{Nuclear Physics}

\begin{document}
\begin{frontmatter}

\title{
Localization and Large N reduction on $S^3$\\
for the Planar and M-theory limit
}
\author[add1,add2]{Masazumi Honda}
\ead{mhonda@post.kek.jp}
\author[add3]{Yutaka Yoshida}
\ead{yyoshida@post.kek.jp}

\address[add1]{Department of Particle and Nuclear Physics,Graduate University for Advanced Studies (SOKENDAI),
Tsukuba, Ibaraki 305-0801, Japan}
\address[add2]{Kavli Institute for Theoretical Physics, University of California,
Santa Barbara, CA 93106-4030, USA}
\address[add3]{High Energy Accelerator Research Organization (KEK),Tsukuba, Ibaraki 305-0801, Japan}



\begin{abstract}
We show a large N reduction on $S^3$ in a BPS sector for a broad class of theories : 
$\mathcal{N}\geq 2$ supersymmetric Chern-Simons theory with any number of adjoint and bi-fundamental chiral multiplets.
We show that a localization method can be applied to the reduced  model and the path integral 
can be written by a multi-contour integral. 
By taking a particular localization configuration, 
we also show that the large N equivalence between the original theory on $S^3$ and the reduced model 
holds for the free energy and the expectation value of  BPS Wilson loops.
It turns out that the large N reduction on $S^3$ holds also for the M-theory limit.
\end{abstract}


\begin{keyword}
Large N reduction\sep M-theory \sep Localization \sep Matrix model 
\end{keyword}

\date{March 2012}

\end{frontmatter}


\tableofcontents

\section{Introduction}
Large $N$ gauge theory is one of key ingredient for exploring non-perturbative aspects of gauge theory and string theory.
For example, the $1/N$ expansion \cite{'tHooft:1973jz} has been an useful tool for understanding the phase diagram of QCD.
Furthermore, the gauge/gravity duality \cite{Maldacena:1997re,Gubser:1998bc,Witten:1998qj} and 
the matrix model \cite{Ishibashi:1996xs,Banks:1996vh,Dijkgraaf:1997vv} have suggested that
many large N gauge theories are related to string theories.
While large $N$ limit often make analysis of gauge theory simpler,
it is generally difficult to solve the large $N$ limit of the gauge theories.

However, a drastic simplification occurs 
by using the large $N$ reduction \cite{Eguchi:1982nm} for some large $N$ gauge theories.
It asserts that the planar large $N$ limit of gauge theories can be studied by their reduced models with some assumptions, 
which can be obtained by dimensional reduction.
The original idea does not work in general because of the
spontaneous breaking of the U(1)$^D$ symmetry in the reduced model \cite{Bhanot:1982sh},
which led to various proposals \cite{Bhanot:1982sh,Parisi:1982gp,Gross:1982at,Das:1982ux,GonzalezArroyo:1982hz,
Narayanan:2003fc,Kovtun:2007py}.
  
A naive question is ``how is the large N reduction generalized to curve space?''.
This generalization was firstly proposed in \cite{Ishii:2008ib} for $S^3$ and
then generalized to the case for semi-simple compact group manifolds \cite{Kawai:2009vb} 
and their coset spaces \cite{Kawai:2010sf}.  
This proposal is based on correspondence of each Feynman diagram and
lifting flat directions up due to mass terms.
If such a generalization is possible, this can give an insight to emergent geometry in matrix model and
non-perturbative regularization of field theories on curved space\footnote{
There are some proposals in this direction \cite{Hanada:2009hd,Hanada:2009kz,Hanada:2010kt,Hanada:2011qx}.
}.
However, sufficient conditions for the correspondence has not been established yet and
there are only few examples of nontrivial test 
\cite{Ishii:2007sy,Ishiki:2010pe,Ishiki:2011ct,Honda:2010nx,Honda:2011qk,Nishimura:2009xm}.

In this paper, we consider the large N reduction on $S^3$ for a broad class of theories: 
three dimensional $\mathcal{N}\geq 2$ supersymmetric quiver Chern-Simons matter theories (CSM) 
with any number of adjoint and bi-fundamental chiral multiplets.
For example, such a class of theory includes the ABJM theory \cite{Aharony:2008ug} as a special case,
which is the leading candidate of low-energy effective theory of M2-branes. 
While many supersymmetric quiver CSM theories in the planar limit have been conjectured to be dual to
superstring theories on certain backgrounds,
the gauge/gravity duality suggests that
these theories are also dual to M-theories on certain backgrounds \cite{Aharony:2008ug,Imamura:2008nn,Jafferis:2008qz}
for another large N limit called ``M-theory limit''.
Although the large N reduction has been considered only for the planar limit so far\footnote{
The so-called orbifold equivalence for the M-theory limit of the ABJM theory \cite{Aharony:2008ug}
was considered in \cite{Hanada:2011yz,Hanada:2011zx}.
},
here we ask a question: ``Does such a drastic simplification occur also for the M-theory limit?''.
This question is highly nontrivial in the following reasons.
First of all, we do not well understand  general properties of the field theory in the M-theory limit  
although there are recently a few developments \cite{Herzog:2010hf,Marino:2011eh}.
Secondly we cannot use any perturbative arguments in the M-theory limit.
Finally, 
it is nontrivial whether an usual large N factorization  as for the planar limit occurs or not in this limit.
Therefore, we expect that 
usual arguments by Schwinger-Dyson equation \cite{Eguchi:1982nm} and coherent state \cite{Yaffe:1981vf,Kovtun:2007py}
are not also useful.
Thus, we need a non-perturbative method in order to answer the question.
In this paper, we adopt a localization method to study non-perturbative aspects of the theories as such a method. 

Localization methods have been played important roles 
and brought many exact analyses in (topologically twisted) supersymmetric theories.
Generically, it is difficult to evaluate path integrals exactly in quantum field theories 
or even in reduced models with finite degrees of freedom such as instanton partition functions. 
When a theory possesses supersymmetry and one can apply a localization formula in the theory, 
the path integral reduces to a multi-contour integral (matrix model) or a summation. 
For example, the path integrals for instanton partition functions 
in four dimensional $\mathcal{N}=2$ supersymmetric gauge theories  
are calculated exactly by an equivariant localization formula  
and reduce to finite summation labeled by Young diagrams \cite{Nekrasov:2002qd}. 
The analyses by the localization formula reproduce the results  obtained 
by analyzing infrared structure of coulomb moduli spaces \cite{Seiberg:1994rs}.

Recently, there have been many progresses in a localization method 
for four dimensional $\mathcal{N} \ge 2 $ rigid supersymmetric field theories on spheres 
attributed to \cite{Pestun:2007rz}.
For instance, 
it is shown  that 
the expectation value of the circular BPS Wilson loop in the $\mathcal{N}=4$ super Yang-Mills (SYM) theory on $S^4$ 
can be described by a Gaussian matrix model. 
This originally has been conjectured in \cite{Erickson:2000af,Drukker:2000rr} 
in the context of $\mathrm{AdS}_5/\mathrm{CFT}_4$ correspondence.

The localization method can be also applied to supersymmetric CSM theories on $S^3$ 
and has expected to be useful for quantitative tests of the $\mathrm{AdS}_4/\mathrm{CFT}_3$ correspondence. 
In fact, the authors of  \cite{Kapustin:2009kz} constructed off-shell $\mathcal{N}=2$ supersymmetry on $S^3$ 
and showed that 
the expectation values of the BPS Wilson loops and the partition functions 
in the $\mathcal{N}=2$ supersymmetric CSM theories  
can be described by certain matrix models. 
This is generalized  to  the  general $R$-charge assignments for matter chiral multiplets 
in \cite{Jafferis:2010un,Hama:2010av}. 
Especially, the ABJM matrix model is analytically continued to the CS matrix model 
on the lens space $S^3/\Z_2$ \cite{Marino:2009jd}. 
Large N-duality between the pure CS theory on the lens space and topological string 
on local $\P^1 \times \P^1$ \cite{Aganagic:2002wv} 
enables to derive large 't Hooft coupling behavior of the BPS Wilson loops \cite{Marino:2009jd} 
and the degrees of freedom of multi-parallel M2-branes\footnote{
The free energy in the ABJM theory is also numerically studied for arbitrary rank and level in \cite{Hanada:2012si}.
} \cite{Drukker:2010nc}.   
Many other application based on localization methods in three dimensional supersymmetric theories are achieved, 
for example see 
\cite{Kapustin:2010xq,Jafferis:2010un,Hama:2010av,Herzog:2010hf,Hama:2011ea,Imamura:2011wg,Jafferis:2011zi}.

In this paper, we consider the large N reduction on $S^3$ 
for any $\mathcal{N}\geq 2$ supersymmetric quiver CSM theories.
In this class of theory,
we show via the localization method  that 
the large N reduction on $S^3$ for a kind of BPS operators holds
both in the planar and M-theory limit.

This article is organized as follows. 
In section~\ref{review}, we briefly review the large N reduction on $S^3$ and
construct the reduced model for general $\mathcal{N}\geq 2$ supersymmetric Chern-Simons theory.
In section~\ref{localization}, we apply the localization method to the reduced model 
and show the partition function is described by matrix models.
In section~\ref{sec:planar} and \ref{sec:M-theory}, 
we argue about the large N correspondence for the planar and M-theory limit, respectively.
Section~\ref{sec:conclusion} is devoted to a conclusion.

\section{Review of large $N$ reduction on $S^3$}
\label{review}
In this section, we review the large $N$ reduction on $S^3$.
For the detail, see \cite{Ishiki:2011ct}.

\subsection{Large N reduction on $S^1$}
In order to give intuitive understanding of the large N reduction for readers,
we first consider the large N reduction on $S^1$ as the simplest example \cite{Kawai:2007tn}. 
Let us consider the matrix quantum mechanics on $S^1$ with the radius $l$, whose action is
\begin{\eq}
S=\frac{1}{g^2} \int_0^{2\pi l} dx
\Tr \left(  \frac{1}{2}\left( \frac{d\phi}{dx} \right)^2 +\frac{\xi^2}{2l^2}\phi^2 +\frac{1}{4}\phi^4 \right) ,
\end{\eq}
where $\phi(x)$ is an $N\times N$ hermitian matrix valued and $\xi$ is the dimensionless mass. 
Making the Fourier transformation $\phi (x) =\sum_{n=-\infty}^\infty \phi^{(n)} e^{inx/R} $,
the action in the momentum representation is given by
\begin{\eqa}
S
&=& \frac{V_{S^1}}{g^2}
\Tr  \Biggl[  \frac{1}{2l^2} \sum_n (n^2 +\xi^2 )\phi^{(n)} \phi^{(-n)}  \NN \\
&& +\frac{1}{4}\sum_{n_1 ,n_2  ,n_3 ,n_4}\delta_{n_1 +n_2 +n_3 +n_4 ,0}\phi^{(n_1)}\phi^{(n_2)}\phi^{(n_3)}\phi^{(n_4)}
   \Biggr] , 
\end{\eqa}
where $V_{S^1}=2\pi l$ is the volume of $S^1$.
Let us consider the free energy in the 't~Hooft limit:
\begin{\eq}
N\rightarrow\infty \quad {\rm with}\ \lambda =g^2 N =\rm{fixed}.
\end{\eq}
The planar contribution at 2-loop level is
\begin{\eqa}
\frac{F_{\rm planar}^{\rm 2-loop}}{V_{S^1}}
&=& \frac{1}{2g^2}\sum_{n_1 ,n_2  ,n_3 ,n_4}\delta_{n_1 +n_2 +n_3 +n_4 ,0}\  
      \overline{\phi^{(n_1)}_{ab} \phi^{(n_2)}_{bc}}\cdot \overline{\phi^{(n_3)}_{cd}\phi^{(n_4)}_{da}}   \NN \\
&=& \frac{1}{2g^2} \left( \frac{g^2 l^2}{V_{S^1}} \right)^2
    \sum_{n_1 ,n_3} \frac{\delta_{ac}\delta_{bb}\delta_{ca}\delta_{dd}}{(n_1^2 +\xi^2 )(n_3^2 +\xi^2 )} \NN \\
&=&  N^2 \cdot \frac{\lambda l^4 }{2V_{S^1}^2} 
    \sum_{n_1 ,n_2} \frac{1}{(n_1^2 +\xi^2 )(n_2^2 +\xi^2 )}, 
\end{\eqa}
where $\overline{\phi^{(n_1)}_{ab} \phi^{(n_2)}_{cd}}$ denotes the propagator.

In order to obtain the reduced model, we apply the following rule:
\begin{\eq}
\phi (x) \rightarrow e^{iPx}\phi e^{-iPx},\quad g \rightarrow g_r ,
\end{\eq}
where $\phi$ is an $M\times M$ constant hermitian matrix and $P$ is the diagonal matrix taking the form
\begin{\eq}
P = \frac{1}{l} {\rm diag} \left(  \frac{-\nu +1}{2},\frac{-\nu +3}{2},\cdots ,
\frac{\nu -1}{2}  \right) \otimes \mathbf{1}_N 
\quad {\rm with}\quad \nu N=M.
\label{S1_momentum}
\end{\eq}
Then the action of the reduced model is
\begin{\eq}
S_r =\frac{V_{S^1}}{g_r^2} 
\Tr_M \left(  -\frac{1}{2} [P,\phi ]^2 +\frac{\xi^2}{2l^2}\phi^2 +\frac{1}{4}\phi^4 \right) ,
\end{\eq}
where $\Tr_M$ stands for the trace over $M\times M$ matrices.
If we decompose $\phi$ into a $N\times N$ matrix $\phi^{(s,t)}\ (s,t =1,2,\cdots ,\nu )$ as
\begin{\eq}
\phi 
=\begin{pmatrix}
\phi^{(1,1)}     & \ldots  & \phi^{(1,\nu )} \cr
\vdots           & \ddots & \vdots \cr
\phi^{(\nu ,1)} & \ldots  & \phi^{(\nu ,\nu )}  
\end{pmatrix}, 
\end{\eq}
then we can rewrite the action as
\begin{\eq}
S_r =\frac{V_{S^1}}{g_r^2} 
\Tr \Biggl[  \frac{1}{2l^2}\sum_{s,t}\left( (P_s -P_t )^2 +\xi^2  \right) \phi^{(s,t)}\phi^{(t,s)}     
              +\frac{1}{4}\sum_{s,t,u,v} \phi^{(s,t)}\phi^{(t,u)}\phi^{(u,v)}\phi^{(v,s)} \Biggr] ,
\end{\eq}
where $P_s =\frac{-\nu +1}{2}+(s-1)$.
Now let us compute the free energy in the reduced model at 2-loop level and
take the 't~Hooft limit:
\begin{\eq}
N\rightarrow\infty ,\quad \nu\rightarrow\infty \quad {\rm with}\ \lambda_r =g_r^2 N =\rm{fixed}.
\end{\eq}
The planar contribution at 2-loop level is
\begin{\eqa}
\frac{F_{\rm r,planar}^{\rm 2-loop}}{V_{S^1}}
&=& \frac{1}{2g_r^2}\sum_{s,t,u,v} 
     \overline{\phi^{(s,t)}_{ab} \phi^{(t,u)}_{bc}}\cdot \overline{\phi^{(u,v)}_{cd}\phi^{(v,s)}_{da}}   \NN \\
&=& \frac{1}{2g_r^2} \left( \frac{g_r^2 l^2}{V_{S^1}} \right)^2
    \sum_{s,t,u,v} \delta_{su}\delta_{tt}\delta_{vv}
    \frac{\delta_{ac}\delta_{bb}\delta_{ca}\delta_{dd}}{((P_s -P_t)^2 +\xi^2 )((P_u -P_v)^2 +\xi^2 )} \NN \\
&=&  N^2  \nu \cdot \frac{\lambda_r l^4 }{2V_{S^1}^2} 
    \sum_{n_1 ,n_2} \frac{1}{(n_1^2 +\xi^2 )(n_2^2 +\xi^2 )}.
\end{\eqa}

Therefore, if we identify $\lambda_r =\lambda$, we find
\begin{\eq}
\frac{F_{\rm planar}^{\rm 2-loop}}{N^2}   = \frac{F_{\rm r,planar}^{\rm 2-loop}}{N^2 \nu}.
\end{\eq}
Although the non-planar diagrams do not correspond with each other,
these are relatively suppressed by the order of $\mathcal{O}(1/N^2 )$ against the planar diagrams.
This correspondence is based on coincidence of all planar diagrams.
Intuitively, the constant matrix $P$ (\ref{S1_momentum}) supplies 
the ``missing Kaluza-Klein momenta'' along the $S^1$-direction associated with the dimensional reduction.
Such a mechanism occurs only for the planar diagrams in general.
From this point of view,
we can regard the role of the parameter $\nu$ as the UV cutoff in the theory.

\subsection{Large N reduction on $S^3$}
In this subsection, we briefly review the large $N$ reduction on $S^3$.
Let us consider the scalar field theory on $S^3$ with the radius $l$, whose action is
\begin{\eq}
S=\frac{V_{S^3}}{g^2} \int \frac{d\Omega_3}{2\pi^2}
\Tr \left(  -\frac{2}{l^2} \left( \mathcal{L}_i \phi \right)^2 +\frac{2\xi^2}{l^2}\phi^2 +\frac{1}{4}\phi^4 \right) ,
\end{\eq}
where $V_{S^3} =2\pi^2 l^3$ is the volume of $S^3$ and $\mathcal{L}_i \ (i=1,2,3)$ is the Killing vector on the unit $S^3$. 
In order to obtain the action in the angular momentum representation, we make the spherical harmonics expansion as
\begin{\eq}
\phi (\Omega_3 ) 
= \sum_J \sum_{m,\tilde{m}=-J}^J \phi_{Jm\tilde{m}} Y_{Jm\tilde{m}}(\Omega_3 )
\equiv \sum_\mathbf{J}  \phi_{\mathbf{J}} Y_{\mathbf{J}}(\Omega_3 ) ,
\end{\eq}
and use the identities
\begin{\eqa}
\mathcal{L}_i^2 Y_{\mathbf{J}}(\Omega_3 )  &=& J(J+1)Y_{\mathbf{J}}(\Omega_3 ), \\
\int \frac{d\Omega_3}{2\pi^2} Y^\ast_{\mathbf{J_1}}(\Omega_3 ) Y_{\mathbf{J_2}}(\Omega_3 )
&=& \delta_{J_1 J_2}\delta_{m_1 m_2}\delta_{\tilde{m}_1 \tilde{m}_2},
\end{\eqa}
where $Y^\ast_{\mathbf{J}}(\Omega_3 ) $ is given by
$Y^\ast_{\mathbf{J}}(\Omega_3 ) \equiv (-1)^{m-\tilde{m}} Y_{\mathbf{J^\ast}}(\Omega_3 ) $ with $\mathbf{J^\ast} =(J,-m,-\tilde{m})$.
Then we can rewrite the action as
\begin{\eqa}
S &=& \frac{V_{S^3}}{g^2} \Tr 
    \Biggr( \frac{2}{l^2} \sum_{\mathbf{J}}(-1)^{m-\tilde{m}} (J(J+1)+\xi^2 ) \phi_{\mathbf{J}}\phi_{\mathbf{J^\ast}} \NN \\
&&         +\frac{1}{4} \sum_{\mathbf{J_1} ,\mathbf{J_2},\mathbf{J_3},\mathbf{J_4}}  
                        V^{(4)}_{\mathbf{J_1}\mathbf{J_2}\mathbf{J_3}\mathbf{J_4}}
                         \phi_{\mathbf{J_1}}\phi_{\mathbf{J_2}}\phi_{\mathbf{J_3}}\phi_{\mathbf{J_4}}  \Biggr) ,
\end{\eqa}
where $V^{(4)}_{\mathbf{J_1}\mathbf{J_2}\mathbf{J_3}\mathbf{J_4}}$ is the 4-point vertex
\begin{\eq}
V^{(4)}_{\mathbf{J_1}\mathbf{J_2}\mathbf{J_3}\mathbf{J_4}}
=\int \frac{d\Omega_3}{2\pi^2} Y_{\mathbf{J_1}}(\Omega_3 )Y_{\mathbf{J_2}}(\Omega_3 )Y_{\mathbf{J_3}}(\Omega_3 )Y_{\mathbf{J_4}}(\Omega_3 ) .
\end{\eq}
Let us consider the free energy 
in the 't~Hooft limit ($N\rightarrow\infty \ {\rm with}\ \lambda =g^2 N =\rm{fixed}$) again.
The planar contribution at 2-loop level is
\begin{\eqa}
\frac{F_{\rm planar}^{\rm 2-loop}}{V_{S^3}}
&=& \frac{1}{2g^2}\sum_{\mathbf{J_1} ,\mathbf{J_2},\mathbf{J_3},\mathbf{J_4}}   
     V^{(4)}_{\mathbf{J_1}\mathbf{J_2}\mathbf{J_3}\mathbf{J_4}}\ 
        \overline{\phi_{\mathbf{J_1},ab} \phi_{\mathbf{J_2},bc}}\cdot 
        \overline{\phi_{\mathbf{J_3},cd}\phi_{\mathbf{J_4},da}}   \NN \\
&=& \frac{1}{2g^2} \left( \frac{g^2 l^2}{4V_{S^3}} \right)^2
       \sum_{\mathbf{J_1},\mathbf{J_3}}   V^{(4)}_{\mathbf{J_1}\mathbf{J_1^\ast} \mathbf{J_3}\mathbf{J_3^\ast}}
         \delta_{ac}\delta_{bb}\delta_{ca}\delta_{dd}
        \frac{(-1)^{m_1 -\tilde{m}_1}}{J_1 (J_1 +1) +\xi^2} \frac{(-1)^{m_3 -\tilde{m}_3}}{J_3 (J_3 +1) +\xi^2} \NN \\
&=& N^2 \cdot \frac{\lambda l^4 }{8V_{S^3}^2}
       \sum_{\mathbf{J_1},\mathbf{J_2}}   V^{(4)}_{\mathbf{J_1}\mathbf{J_1^\ast} \mathbf{J_2}\mathbf{J_2^\ast}}
        \frac{(-1)^{m_1 -\tilde{m}_1}}{J_1 (J_1 +1) +\xi^2} \frac{(-1)^{m_2 -\tilde{m}_2}}{J_2 (J_2 +1) +\xi^2}.
\end{\eqa}

In order to obtain the reduced model, we apply the following rule:
\begin{\eq}
\phi (\Omega_3 ) \rightarrow G^{-1} \phi G,\quad g \rightarrow g_r,
\label{rule_S3}
\end{\eq}
where $\phi$ is an $M\times M$ constant hermitian matrix again 
and $G$ is the representation matrix of $SU(2)$ in the $M$-dimensional representation whose generator is\footnote{
In \cite{Kawai:2009vb}, the authors have been proposed another representation of the generator
$L_i = \bigoplus_{s=1}^{\nu} L_i^{(s)} \otimes \mathbf{1}_{s} \otimes \mathbf{1}_{N}$.
Here we do not consider the background.}
\begin{\eq}
L_i = \bigoplus_{s=1}^{\nu} L_i^{(n_s)} \otimes \mathbf{1}_{N} \quad {\rm with} \quad n_s =n+s-\frac{\nu +1}{2},
\end{\eq}
where $L_i^{(n)}$ denotes the $n$-dimensional irreducible representation of $SU(2)$.

Then the action of the reduced model is
\begin{\eq}
S_r =\frac{V_{S^3}}{g_r^2} 
\Tr_M \left(  -\frac{2}{l^2} [L_i ,\phi ]^2 +\frac{\xi^2}{2l^2}\phi^2 +\frac{1}{4}\phi^4 \right) .
\end{\eq}
If we decompose $\phi$ into a $n_s N\times n_t N$ matrix $\phi^{(s,t)}\ (s,t =1,2,\cdots ,\nu )$ again as
\begin{\eq}
\phi 
=\begin{pmatrix}
\phi^{(1,1)}     & \ldots  & \phi^{(1,\nu )} \cr
\vdots           & \ddots & \vdots \cr
\phi^{(\nu ,1)} & \ldots  & \phi^{(\nu ,\nu )}  
\end{pmatrix}, 
\end{\eq}
and expand $\phi^{(s,t)}$ in terms of the scalar fuzzy sphere harmonics\footnote{
For details of the fuzzy sphere harmonics, see \ref{sec:fuzzy}.}:
\begin{\eq}
\phi^{(s,t)} 
= \sum_{J=|j_s -j_t|}^{j_s +j_t} \sum_{m=-J}^J \phi_{Jm}^{(s,t)} \otimes \hat{Y}_{Jm(j_s j_t )}
\equiv \sum_{\bbJ^{st}} \phi_{\bbJ}^{(s,t)} \otimes \hat{Y}_{\bbJ^{st}},
\end{\eq}
where $\phi_{Jm}^{(s,t)}$ is a $N\times N$ matrix.
In order to rewrite the action in a convenient form,
we use the identity
\begin{\eq}
\left( L_i \circ \right)^2 \hat{Y}_{\bbJ^{st}}  =J(J+1) \hat{Y}_{\bbJ^{st}}, 
\end{\eq}
and the orthogonal relation
\begin{\eq}
\frac{1}{n} \tr  \left( \hat{Y}_{\bbJ^{st}_1}^\dag \hat{Y}_{\bbJ^{st}_2} \right)
=\delta_{J_1 J_2}\delta_{m_1 m_2},
\end{\eq}
where $\hat{Y}^\dag_{\bbJ^{st}}$ is given by $\hat{Y}^\dag_{\bbJ^{st}} = (-1)^{m-(j_s -j_t )} Y_{\bbJ^\dag}$ 
with $\bbJ^\dag =(J, -m, j_t ,j_s)$
and $\tr$ stands for the trace over $(2j_t +1)\times (2j_t +1)$ matrices.
Then we can rewrite the action as
\begin{\eqa}
S_r 
&=& \frac{V_{S^3}n}{g_r^2}
\Biggl[ \frac{2}{l^2} \sum_{s,t}\sum_{\bbJ}(-1)^{m-(j_s -j_t )} (J(J+1)+\xi^2 ) \Tr \left( \phi_{\bbJ}^{(s,t)} \phi_{\bbJ^{\dag}}^{(t,s)} \right)  \NN\\
&&         +\frac{1}{4} \sum_{s,t,u,v}\sum_{\bbJ_1^{st} ,\bbJ_2^{tu} ,\bbJ_3^{uv} ,\bbJ_4^{vs} } 
                          \hat{V}^{(4)}_{\bbJ_1^{st}\bbJ_2^{tu}\bbJ_3^{uv}\bbJ_4^{vs}}
                       \Tr \left( \phi_{\bbJ_1}^{(s,t)} \phi_{\bbJ_2}^{(t,u)}\phi_{\bbJ_3}^{(u,v)}\phi_{\bbJ_4}^{(v,s)} \right)
\Biggr],
\end{\eqa}
where $\hat{V}^{(4)}_{\bbJ_1^{st}\bbJ_2^{tu}\bbJ_3^{uv}\bbJ_4^{vs}}$ is the 4-point vertex
\begin{\eq}
\hat{V}^{(4)}_{\bbJ_1^{st}\bbJ_2^{tu}\bbJ_3^{uv}\bbJ_4^{vs}}
=\frac{1}{n}\tr \left( \hat{Y}_{\bbJ^{st}_1} \hat{Y}_{\bbJ^{tu}_2} \hat{Y}_{\bbJ^{uv}_3} \hat{Y}_{\bbJ^{vs}_4}  \right) .
\end{\eq}

Let us consider the free energy in the following limit:
\begin{\eq}
N\rightarrow\infty ,\quad \nu\rightarrow\infty ,\quad \frac{n}{\nu}\rightarrow \infty
\quad {\rm with}\ \lambda_r =\frac{g_r^2 N}{n} =\rm{fixed},
\label{limit_S3}
\end{\eq}
which is the counter part of the 't~Hooft limit in the original theory.

The planar contribution at 2-loop level is
\begin{\eqa}
\frac{F_{\rm r,planar}^{\rm 2-loop}}{V_{S^3}}
&=& \frac{n}{2g_r^2}\sum_{s,t,u,v} \sum_{\bbJ_1^{st} ,\bbJ_2^{tu} ,\bbJ_3^{uv} ,\bbJ_4^{vs} }  
      \hat{V}^{(4)}_{\bbJ_1^{st}\bbJ_2^{tu}\bbJ_3^{uv}\bbJ_4^{vs}}\ 
          \overline{\phi_{\bbJ_1 ,ab}^{(s,t)} \phi_{\bbJ_2 ,bc}^{(t,u)} }\cdot 
          \overline{\phi_{\bbJ_3 ,cd}^{(u,v)} \phi_{\bbJ_4 ,da}^{(v,s)} }   \NN \\
&=& \frac{n}{2g_r^2} \left( \frac{g_r^2 l^2}{4nV_{S^3}} \right)^2 \sum_{s,t,v} \sum_{\bbJ_1^{st} ,\bbJ_3^{sv}  }  
      \hat{V}^{(4)}_{\bbJ_1^{st}\bbJ_1^{ts\dag}\bbJ_3^{sv}\bbJ_3^{vs\dag}} 
         \delta_{ac}\delta_{bb}\delta_{ca}\delta_{dd}
        \frac{(-1)^{m_1 -(j_s -j_t )}}{J_1 (J_1 +1) +\xi^2} \frac{(-1)^{m_3 -(j_s -j_v )}}{J_3 (J_3 +1) +\xi^2} \NN \\
&=& N^2 \cdot \frac{\lambda_r l^4 }{8V_{S^3}^2} \sum_{s,t,v} \sum_{\bbJ_1^{st} ,\bbJ_2^{sv}  }  
      \hat{V}^{(4)}_{\bbJ_1^{st}\bbJ_1^{ts\dag}\bbJ_2^{sv}\bbJ_2^{vs\dag}} 
        \frac{(-1)^{m_1 -(j_s -j_t )}}{J_1 (J_1 +1) +\xi^2} \frac{(-1)^{m_2 -(j_s -j_v )}}{J_2 (J_2 +1) +\xi^2}.
\end{\eqa}
In the limit (\ref{limit_S3}), 
we can identify as $\tilde{m}_1 =j_s -j_t ,\tilde{m}_2 =j_s -j_v ,\tilde{m}_3 =j_t -j_v $ 
by reading from the relation \cite{Ishii:2008ib}
\begin{\eq}
\hat{V}^{(4)}_{\bbJ_1^{st}\bbJ_1^{ts\dag}\bbJ_2^{sv}\bbJ_2^{vs\dag}} \quad \rightarrow \quad
V^{(4)}_{\mathbf{J_1}\mathbf{J_1^\ast} \mathbf{J_2}\mathbf{J_2^\ast}}.
\end{\eq}
Therefore, if we make the identification $\lambda_r =\lambda$, we obtain
\begin{\eqa}
\frac{F_{\rm r,planar}^{\rm 2-loop}}{N^2 \nu V_{S^3}}
&\rightarrow & 
\frac{1}{\nu} \cdot \frac{\lambda_r l^4 }{8V_{S^3}^2} \sum_{\tilde{m}_3} \sum_{\bfJ_1 ,\bfJ_2  }  
V^{(4)}_{\mathbf{J_1}\mathbf{J_1^\ast} \mathbf{J_2}\mathbf{J_2^\ast}}
        \frac{(-1)^{m_1 -(j_s -j_t )}}{J_1 (J_1 +1) +\xi^2} \frac{(-1)^{m_2 -(j_s -j_v )}}{J_2 (J_2 +1) +\xi^2} \NN \\
&=& \frac{\lambda_r l^4 }{8V_{S^3}^2} \sum_{\tilde{m}_3} \sum_{\bfJ_1 ,\bfJ_2  }  
V^{(4)}_{\mathbf{J_1}\mathbf{J_1^\ast} \mathbf{J_2}\mathbf{J_2^\ast}}
        \frac{(-1)^{m_1 -(j_s -j_t )}}{J_1 (J_1 +1) +\xi^2} \frac{(-1)^{m_2 -(j_s -j_v )}}{J_2 (J_2 +1) +\xi^2} \NN \\
&=& \frac{F_{\rm planar}^{\rm 2-loop}}{N^2 V_{S^3}}.
\end{\eqa}

Although the case for $S^3$ seems more complicated than the case for $S^1$,
essential features are same with each other.
Similarly for the $S^1$ case,
the parameters $n$ and $\nu$ play the role as the UV cutoff of the angular momentum along $S^3$\footnote{
Strictly speaking, when we regard $S^3$ as the $S^1$ bundle over $S^2$,
the parameter $n$ and $\nu$ correspond to the UV cutoff along $S^2$ and $S^1$, respectively \cite{Ishii:2008ib}.
}.
While we have  demonstrated large N equivalence at two-loop level only for the free energy,
we can also see perturbative coincidence for correlation functions \cite{Ishiki:2011ct}.

\subsection{Construction of Large N reduced model for supersymmetric quiver CSM on $S^3$}
In this subsection, we construct large N reduced models for supersymmetric quiver CSM theories 
on $S^3$ \cite{Hanada:2009hd}.
In general, the action of supersymmetric quiver CSM theory is decomposed as 
\begin{\eq}
S=S_{\rm CS} +S_{\rm YM} +S_{\rm matter},
\end{\eq}
where $S_{\rm CS}$, $S_{\rm YM}$ and $S_{\rm matter}$ are 
the action of the Chern-Simons, Yang-Mills and matter part, respectively.
In the following, we construct the reduced model of each part.

\subsubsection*{CS part}
The Chern-Simons action for the $\mathcal{N}=2$ vector multiplet  is given by
\begin{equation}
S_{CS} 
= -\frac{ik}{4\pi} \int \Tr\Bigl[ A\wedge dA -\frac{2}{3} i A\wedge A\wedge A+( -\bar{\lambda}\lambda +2\sigma D) \sqrt{g} d^3 x \Bigr],
\label{original_CS}
\end{equation}
where $k$ is the Chern-Simons level.
In order to apply the prescription (\ref{rule_S3}), 
we rewrite the derivative term in terms of the Killing vector.
Expanding the gauge field as $A=X^i e^i$ and using the Meurer-Cartan equation, we derive
\begin{\eqa} 
dA 
&=& dX_i \wedge e^i +X_i de^i \NN \\
&=& i\frac{2}{l} (J_j X_i )  e^j \wedge e^i +\frac{1}{l}\epsilon_{ijk} X_i e^j\wedge e^k \NN \\
&=& \frac{2}{l} \left( i J_i X_j +\frac{1}{2}\epsilon_{ijk}X_k \right) e^i\wedge e^j  ,  
\end{\eqa}
where $J_i$ is the Killing vector on the unit sphere.
In this way, we obtain
\begin{\eqa} 
A\wedge dA -\frac{2}{3} i A\wedge A\wedge A 
&=& \left\{  \frac{2}{l} X_i \left( i J_j X_k +\frac{1}{2}\epsilon_{jkl}X_l \right) -i\frac{2}{3}X_i X_j X_k \right\} 
     e^i \wedge e^j \wedge e^k \NN \\
&=& \left\{  i \frac{2}{l}\epsilon_{ijk} X_i J_j X_k 
            +\frac{2}{l}X_i^2 -i\frac{2}{3}\epsilon_{ijk }X_i X_j X_k \right\} l^3 d\Omega_3 .
\end{\eqa}
The prescription for constructing the reduced model is
\begin{\eq}
\Phi (\Omega_3 ) \ \RA\ G^{-1} \Phi G,\quad 
\frac{k}{4\pi}\ \RA\  \frac{1}{g_{CS,r}^2},
\end{\eq}
where $\Phi$ represents the collection of the components fields in the theory. 
Then, the action of the reduced model is
\begin{equation}
S_{\rm CS}^r
= -\frac{iV_{S^3}}{g_{\rm CS,r}^2}  
 \Tr\Bigl[\  i \frac{2}{l}\epsilon_{ijk} X_i [ L_j ,X_k ] +\frac{2}{l}X_i^2 -i\frac{2}{3}\epsilon_{ijk }X_i X_j X_k
          -\bar{\lambda}\lambda +2\sigma D \ \Bigr] .
\label{naive_red}
\end{equation}
Here note that we can absorb $-\frac{2}{l}L_i$ into $X_i$ as
\begin{\eq}
-\frac{2}{l}L_i +X_i \ \RA\ X_i .
\end{\eq}
After the absorbing this factor, we obtain the simpler action as
\begin{equation}\label{redcs}
S_{\rm CS}^r
= -\frac{iV_{S^3}}{g_{\rm CS,r}^2}  \Tr\Bigl[\ \frac{2}{l}X_i^2 -i\frac{2}{3}\epsilon_{ijk }X_i X_j X_k
                                          -\bar{\lambda}\lambda +2\sigma D \ \Bigr],
\end{equation}
which corresponds to the dimensional reduction of the original action (\ref{original_CS}).
The equation of motion for $X_i$ is
\begin{\eq}
[\ X_i ,X_j \ ] = -i\frac{2}{l}\epsilon_{ijk}X_k ,
\end{\eq}
which can be solved as
\begin{equation}\label{classical}
X_i = -\frac{2}{l} \bigoplus_{I=1}^{\nu} L_i^{(n_I)} \otimes \mathbf{1}_{N_I}, 
\end{equation}
where $L_i^{(n_I)}$ denotes the $n_I$-dimensional irreducible representation of $SU(2)$ with $\sum_I n_I N_I =M$. 
Since this solution includes $-\frac{2}{l}L_i$ as the special case, 
we can realize the action (\ref{naive_red}) if we expand $X_i$ around the solution as
\begin{\eq}
X_i\ \RA\ -\frac{2}{l}L_i +X_i  .
\end{\eq}
Thus, the rule for generating the reduced model of the gauge theory on $S^3$ becomes simpler as follows
\begin{\eq}
\Phi (\Omega_3 )\ \RA\  \Phi  ,\quad 
J_i \Phi (\Omega_3 )\ \RA\ 0,\quad 
\frac{k}{4\pi}\ \RA\  \frac{1}{g_{\rm CS,r}^2}.
\end{\eq}
In order to realize the original theory, we have to expand
the gauge field as $X_i\ \RA\ -\frac{2}{l}L_i +X_i$.

\subsubsection*{YM part}
The action of the $\mathcal{N}=2$ SYM on $S^3$ is given by
\begin{eqnarray}
S_{\rm YM}
&=& \frac{V_{S^3}}{g_{\rm YM}^2} \int \frac{d\Omega_3}{2\pi^2}\ 
\Tr\Bigl[ \frac{1}{4}F_{ij}^2 +\frac{1}{2}(D_i\sigma )^2 +\frac{1}{2}\left( D+\frac{\sigma}{l} \right)^2 \nonumber \\
&&\ \ \ \ \ \ \ \ \ \ \ 
   +\frac{i}{2}\bar{\lambda}\gamma^i D_i \lambda 
   +\frac{i}{2}\bar{\lambda}[\ \sigma ,\lambda\ ] -\frac{1}{4l}\bar{\lambda}\lambda \Bigr],
\end{eqnarray}
where $D_i \sigma = \partial_i \sigma -i [\ A_i ,\sigma\ ] $. 
Similarly for the Chern-Simons term, we rewrite the field strength as 
\begin{eqnarray*}
F 
&=& dA -iA\wedge A \\
&=& \frac{1}{2}\epsilon_{ijk} \left\{ \frac{2}{l} i\epsilon_{klm}J_l X_m +\frac{2}{l}X_k 
                                      -\frac{i}{2}\epsilon_{klm} [\ X_l ,X_m\ ] \right\}    e^i \wedge e^j.
\end{eqnarray*}
The covariant derivative of the fermion is
\begin{eqnarray*}
\gamma^i D_i \lambda
&=& \gamma^a e_a^i ( \partial_i +\frac{1}{4}\omega_{i}^{a^\prime b^\prime}\lambda -i\gamma^i [\ A_i ,\lambda\ ] \\
&=& \frac{2i}{l} \gamma^a J^a \lambda +\frac{3i}{2l}\lambda -i\gamma^i [\ A_i ,\lambda\ ]
\end{eqnarray*}

Applying the rule
\begin{\eq}
\Phi (\Omega_3 )\ \RA\  \Phi  ,\quad 
J_i \Phi (\Omega_3 )\ \RA\ 0,\quad 
g_{\rm YM} \ \RA\  g_{\rm YM,r},
\end{\eq}
the action of the reduced model is
\begin{eqnarray}\label{qexact}
S_{\rm YM}^r
&=& \frac{V_{S^3}}{g_{\rm YM,r}^2}  
\Tr \Bigl[\ \frac{1}{2} \left(  \frac{2}{l}X_i -\frac{i}{2}\epsilon_{ijk} [\ X_j ,X_k\ ] \right)^2
         -\frac{1}{2} [\ X_i ,\sigma\ ]^2 +\frac{1}{2}\left( D+\frac{\sigma}{l} \right)^2 \nonumber \\
&&\ \ \ \ \ \ \ \ \ \ \ 
   +\frac{1}{2}\bar{\lambda}\gamma^i [\ X_i ,\lambda\ ] +\frac{i}{2}\bar{\lambda}[\ \sigma ,\lambda\ ] 
   -\frac{1}{l}\bar{\lambda}\lambda\ \Bigr].
\end{eqnarray}

\subsubsection*{Matter sector (bi-fundamental)}
The action of the chiral multiplet with the bi-fundamental representation\footnote{
Note that this case also includes an adjoint matter representation under $G=U(N)$ as a special case.
} under the gauge group $G=U(N_1 )\times U(N_2 )$ is given by
\begin{equation}
S_{\rm matter}= l^3 \int d\Omega_3 \ \left( \mathcal{L}_{\rm kin} +\mathcal{L}_{\rm pt} \right) ,
\end{equation}
where $\mathcal{L}_{\rm kin}$ and $\mathcal{L}_{\rm pt}$ are a kinetic term 
and a potential term with higher powers of the matter fields, respectively.
$\mathcal{L}_{\rm kin}$ is given by\footnote{Since $\mathcal{L}_{\rm pt}$ is irrelevant in the context of this paper,
we do not write down this explicitly.}
\begin{eqnarray}
\mathcal{L}_{\rm kin} 
&=&  \Tr \Biggl[
     D_i \bar{\phi} D^i \phi +\bar{\phi}( \sigma_A -\sigma_B )^2 \phi 
    +\frac{i(2q-1)}{l} \bar{\phi}(\sigma_A -\sigma_B ) \phi \nonumber \\
&&  +\frac{q(2-q)}{l^2}\bar{\phi}\phi  +i\bar{\phi}(D_A -D_B ) \phi  +\bar{F}F \nonumber \\
&&  -i\bar{\psi}\gamma^i D_i \psi +i\bar{\psi}(\sigma_A -\sigma_B )  \psi -\frac{2q-1}{2l}\bar{\psi}\psi 
     +i\bar{\psi}(\lambda_A -\lambda_B ) \phi -i\bar{\phi}( \bar{\lambda}_A -\bar{\lambda}_B )\psi  
    \Biggr], \nonumber \\
\end{eqnarray}
where $q$ is the dimension and R-charge of $\phi$.
Each matter field is $N_1 \times N_2$ matrix and $D_i \phi = \partial_i \phi -iA_i\phi +i\phi B_i $.

The reduced model is given by
\begin{eqnarray}\label{redmatter}
\mathcal{L}_{\rm kin}^{\rm r} 
&=& \Tr\Biggl[
     ( X_i \phi -\phi Y_i )^\dag  ( X_i {\phi} -\phi Y_i )
     +\bar{\phi}( \sigma_A -\sigma_B )^2 \phi 
    +\frac{i(2q-1)}{l} \bar{\phi}(\sigma_A -\sigma_B ) \phi \nonumber \\ 
&&    +\frac{q(2-q)}{l^2}\bar{\phi}\phi   +i\bar{\phi}(D_A -D_B ) \phi  +\bar{F}F \nonumber \\
&&  -\bar{\psi}\gamma^i (X_i \psi -\psi Y_i ) +i\bar{\psi}(\sigma_A -\sigma_B )  \psi -\frac{q-2}{l}\bar{\psi}\psi 
     +i\bar{\psi}(\lambda_A -\lambda_B ) \phi -i\bar{\phi}( \bar{\lambda}_A -\bar{\lambda}_B )\psi 
    \Biggr], \nonumber \\
\end{eqnarray}
where $(X_i ,\sigma_A ,D_A ,\lambda_A ):M_1 \times M_1$ matrix , $(Y_i ,\sigma_B ,D_B ,\lambda_B ):M_2 \times M_2$ matrix 
and $(\phi ,F,\psi ) :M_1 \times M_2$ matrix.

\section{Localization}
\label{localization}
We briefly  explain the concept of the localization formula in our interest.
The partition function of a supersymmetric theory is written in schematic way as 
\bea
Z=\int \mathcal{D} \Phi \exp( -S[\Phi ] ) ,
\eea
where $\Phi$ represents the collection of the components fields in the theory. 
$S[\Phi ]$ is the action invariant under the nilpotent supercharge $Q$.
We choose one of the supercharges $Q$ and  deform the action by one parameter family of $Q$-exact term 
as $S[\Phi ] + t Q \cdot V[\Phi]$. 
We assume  $Q \cdot V$ respects the symmetry of the theory and its bosonic part is positively semi-definite.
The $Q$-invariance requires that the expectation value of $Q$-closed operator $\mathcal{O}(\Phi)$ 
and th partition function are independent of the coupling parameter $t$. 
When we take the limit $t \to \infty$, the path integral is exactly evaluated 
with the quadratic order of fluctuation fields, 
namely one-loop of $Q \cdot V[\Phi]$ around the localization field configurations $Q \cdot V(\Phi_0)=0$.
Then, the localization configuration only contributes to the action at classical level.   
Then the expectation value is formally written as 
\bea
\langle \mathcal{O} \rangle
= \frac{ \sum_{\Phi_0} \mathcal{O} (\Phi_0) \exp(-S[\Phi_0]) Z_{1-\mathrm{loop}}(\Phi_0)}
       { \sum_{\Phi_0}  \exp(-S[\Phi_0]) Z_{1-\mathrm{loop}}(\Phi_0)}.  
\eea 
Here $\sum_{\Phi_0}$ stands for the summation over the configurations with  $Q \cdot V(\Phi_0)=0$. 
As we will see later, 
it is actually not summation rather multi-contour integrals of the scalar $\sigma$ in the vector multiplet. 
$Z_{1-\mathrm{loop}}$ is the one-loop determinant of the fluctuations around the localization configurations.
In this section, 
we evaluate the one-loop determinant of the dimensional reduced $\mathcal{N}=2$ supersymmetric CSM theories. 

\subsection{Gauge sector}

\subsubsection*{Localized configuration}
Since the reduced $\mathcal{N}=2$ SYM action $S_{YM}^r$ itself is rewritten as 
(For the derivation, see \ref{sec:SUSY})
\begin{equation}
\bar{\epsilon}\epsilon S_{YM}^r 
=\delta_{\bar{\epsilon}}\delta_\epsilon \Tr_M \Bigl[ \frac{1}{2}\bar{\lambda}\lambda -2D\sigma  \Bigr],
\end{equation}
we can choose the deformation term $Q \cdot V$ as the reduced $\mathcal{N}=2$ YM action $S_{YM}^r$.   
From the action (\ref{qexact}), the localized configuration is determined by the equation
\begin{equation}
[\ X_i , X_j\ ] = -i\frac{2}{l} \epsilon_{ijk}X_k, \quad [\ X_i , \sigma\ ]=0,\quad D+\frac{\sigma}{l}=0,
\quad \lambda =\bar{\lambda}=0.
\end{equation}
This can be solved as
\begin{equation}\label{classical}
X_i = -\frac{2}{l} \bigoplus_{s=1}^{\nu} L_i^{(n_s)} \otimes \mathbf{1}_{N_s},\quad 
\sigma =\bigoplus_{s=1}^{\nu} \mathbf{1}_{n_s} \otimes \sigma_{N_s} (:=\bar{\sigma}),\quad 
D=-\frac{\bar{\sigma}}{l}
\end{equation}
where $L_i^{(n_s)}$'s denote the $n_s$-dimensional irreducible representation of $SU(2)$ with $\sum_s n_s N_s =M$. 
Note that there is an important difference from the original theory.
There exists the nontrivial configuration of the gauge field at the localization points
contrary to the case for the original theory.

In order to realize the field theory on $S^3$, 
we specify the representation as $n_s =n+s-\frac{\nu +1}{2},\ N_s =N$, namely, $X_i =-\frac{2}{l}L_i$
and $\sigma_{N_s}=\sigma_0$.
Substituting (\ref{classical}) into the reduced $\mN=2$ supersymmetric Chern-Simons action (\ref{redcs}), 
the CS action on the localized field configurations becomes  
\bea\label{localcs}
S^r_{CS} = \frac{iV_{S^3} n \nu}{g_{CS,r}^2}  \Tr \sigma^2_0 ,
\eea
up to an irrelevant constant.
Here the trace ``$\Tr$'' is taken over $N \times N$ matrix.
 
\subsubsection*{1-loop determinant}
Here, we evaluate the one-loop determinant of the $\mN=2$ SYM action around the localization points (\ref{classical}).
In order to obtain the action at the quadratic order of fluctuation fields, 
we expand the fields around the localized configuration as 
\bea\label{localym}
&&X_i \rightarrow -\frac{2}{l}L_i +\frac{1}{\sqrt{t}} \tilde{X}_i, \quad 
\sigma \rightarrow \bar{\sigma} +\frac{1}{\sqrt{t}} \tilde{\sigma}, \nonumber \\ 
&&D \rightarrow -\frac{1}{l} \sigma_0 +\frac{1}{\sqrt{t}} \tilde{D}, \quad 
\lambda \to \frac{1}{\sqrt{t}}\lambda .
\eea
Then, the quadratic action for the reduced $\mN=2$ SYM is given by
\begin{eqnarray}\label{YM}
\left. S_{\rm YM}^r \right|_{\rm Gauss} 
&=&  
\Tr_M \Bigl[ 
 \frac{1}{2} \left( \frac{2}{l}\right)^2 \left( \tilde{X}_i +i \epsilon_{ijk} [\ L_j , \tX_k\ ]  \right)^2
  -\frac{1}{2} \left( -\frac{2}{l} [\ L_i , \tsig \ ]  + [\ \tX_i , \sigma_0\ ] \right)^2 \nonumber \\
&&  \qquad    -\frac{1}{l} \bar{\lambda}\gamma^i [\ L_i ,\lambda\ ] 
       +\frac{i}{2}\bar{\lambda} [\ \bar{\sigma} ,\lambda\ ] -\frac{1}{l}\bar{\lambda}\lambda + \tilde{D}^2
   \Bigr] .
\end{eqnarray}
Since $\tilde{D}$ has the Gaussian form, this is trivially integrated out. 
In order to perform  the path integral over the fluctuation fields, 
we introduce the vector, scalar and spinor fuzzy sphere harmonics:
$\hat{Y}_{Jm(j_s j_t)i}^{\rho} $, $\hat{Y}_{Jm(j_s j_t)}$ and  $\hat{Y}_{Jm(j_s j_t)\alpha}^{\kappa}$, respectively.
Then the field are  expanded as follows;  
\begin{eqnarray}\label{actionym}
\tilde{X}_i  &=& \sum_{s, t} \sum_{\rho =-1}^1 \sum_{\tilde{Q}=|j_s -j_t|}^{j_s +j_t} \sum_{m=-Q}^Q 
                 \hat{Y}_{Jm(j_s j_t)i}^{\rho} \otimes X_{Jm\rho}^{(s,t)}, \nonumber \\           
\tilde{\sigma} &=&   \sum_{s, t}        \sum_{J=|j_s -j_t|}^{j_s +j_t} \sum_{m=-J}^J 
                   \hat{Y}_{Jm(j_s j_t)} \otimes  \sigma_{Jm}^{(s,t)},  \nonumber \\           
\lambda_{\alpha}  
              &=&  \sum_{s, t} \sum_{\kappa =\pm} \sum_{\tilde{U}=|j_s -j_t|}^{j_s +j_t} \sum_{m=-U}^U 
                 \hat{Y}_{Jm(j_s j_t)\alpha}^{\kappa} \otimes \lambda_{Jm\kappa}^{(s,t)},    
\end{eqnarray}
where $Q=J+\frac{(1+\rho )\rho}{2},\tilde{Q}=J-\frac{(1-\rho )\rho}{2},U=J+\frac{1+\kappa}{4}$ 
and $\tilde{U}=J+\frac{1-\kappa}{4}$\footnote{
More explicitly, these are given by
\begin{eqnarray*}
&& Q|_{\rho =+1}=J+1,\quad \tilde{Q}|_{\rho =+1}=J,\quad  
Q|_{\rho =-1}=J  ,\quad \tilde{Q}|_{\rho =+1}=J+1,\\
&& U|_{\kappa =+1}=J+\frac{1}{2},\quad \tilde{U}|_{\kappa =+1}=J,\quad  
U|_{\kappa =-1}=J            ,\quad \tilde{U}|_{\kappa =+1}=J+\frac{1}{2}.
\end{eqnarray*}
}.
From the properties of the fuzzy sphere harmonics  (\ref{commu}) in \ref{sec:fuzzy},
the quadratic action can be rewritten as
\begin{eqnarray}
&& \left. S_{\rm YM}^r \right|_{\rm Gauss} \NN \\
&=& \sum_{\rho ,\tilde{Q},m} \rho^2 (J+1)^2\ \Tr\left( X_{Jm\rho}^{(s,t)\dag}X_{Jm\rho}^{(s,t)} \right)  
   -\sum_{\rho ,\tilde{Q},m} 
              \Tr\Bigl[\   [\ \sigma_0 , X_{Jm\rho} \ ]^{(s,t)\dag} [\ \sigma_0 ,X_{Jm\rho} \ ]^{(s,t)} \ \Bigr] \NN \\
&&  -\left( \frac{2}{l} \right)^2 \sum_{J,m} J(J+1) \Tr\Bigl[\  \sig_{Jm}^{(s,t)\dag} \sig_{Jm}^{(s,t)} \ \Bigr]     
              -\frac{2}{l}\sum_{J,m}\sqrt{J(J+1)}  
    \Tr\Bigl[\  \sig_{Jm}^{(s,t)\dag} [\  \sigma_0 , X_{Jm0} \ ]^{(s,t)} \ \Bigr] . \nonumber \\ 
\end{eqnarray} 
Next, we introduce the Cartan-Weyl basis  $(H_i ,E_\alpha ,E_{-\alpha})$ satisfying the relations:
\begin{eqnarray}
&& [\ H_i , H_j\ ]=0,\quad [\ H_i , E_\alpha\ ]=\alpha_i \cdot E_\alpha ,\quad 
  [\ E_\alpha , E_{-\alpha}\ ]=\frac{2}{|\alpha |^2} \alpha_i H_i \nonumber \\
&& E_\alpha^\dag =E_{-\alpha},\quad \Tr( E_\alpha E_\beta )=\delta_{\alpha +\beta ,0} .
\end{eqnarray}
Then, we expand each $N\times N$ matrix $X$ in terms of the Cartan-Weyl basis as
\bea
X=X_i H_i +\sum_{\alpha\in\Delta_+} (X_\alpha E_\alpha +X_{-\alpha}E_{-\alpha} )
\eea
and we choose the gauge for the localization configuration $\sigma_0$ 
as $\sigma_0=\mathrm{diag}(\sigma_1,\cdots, \sigma_N)$
\footnote{
For  $U(N)$ case, $\sum_{\alpha \in \Delta} f(\alpha \cdot \sigma)=\sum_{1 \le a \neq b \le N} f(\sigma_a- \sigma_b)$. 
}.
In terms of the basis, the action becomes\footnote{
Here we drop the terms independent of $\sigma$ since these terms become only irrelevant overall constant. }
\begin{eqnarray}
\left. S_{\rm YM}^r \right|_{\rm Gauss} 
&= &    \sum_{J,m,\alpha} \Tr\Biggl[\  \left( \sig_{Jm}^{\alpha (s,t)\dag} , X_{Jm0}^{\alpha (s,t)\dag} \right)
                          \mathcal{C}_J    \begin{pmatrix}\sig_{Jm}^{\alpha (s,t)} \\ X_{Jm0}^{\alpha (s,t)}  \end{pmatrix}
                  \ \Biggr] \nonumber \\
&&    +\sum_{\rho =\pm 1}\sum_{\tilde{Q},m,\alpha} ((J+1)^2+ (\alpha\cdot\sigma )^2 )
              \Tr\Bigl[\ X_{Jm\rho}^{\alpha (s,t)\dag} X_{Jm\rho}^{\alpha (s,t)} \ \Bigr], 
\end{eqnarray}
where $\mathcal{C}_J$ is the $2 \times 2$ kinematic  matrix of $ X_{Jm0}^{\alpha (s,t)}$, $\sig_{Jm}^{\alpha (s,t)} $,
whose component is
\bea\label{kmatrix}
\mathcal{C}_J =
\begin{pmatrix} 
                             \left(\frac{2}{l}\right)^2 J(J+1) & 
                                   \frac{2}{l} \sqrt{J(J+1)} (\alpha\cdot\sigma )\\
                                   \frac{2}{l} \sqrt{J(J+1)} (\alpha\cdot\sigma )  & 
                                                 (\alpha\cdot\sigma )^2 
                              \end{pmatrix}. 
\eea
In order to find the eigenvalues, 
we have to diagonalize the matrix $\mathcal{C}_J$.  
After the straightforward calculation, 
we find that the eigenvalues of the matrix  are $0$ and $ \left( \frac{2}{l}\right)^2 J(J+1) +(\alpha\cdot\sigma )^2$. 
Since the eigenmodes  $(\alpha\cdot\sigma )\sig^{(s,t)} +\frac{2}{l}\sqrt{J(J+1)} X_{Jm0}^{(s,t)}$ 
associated to  zero eigenvalue are  gauge modes, it can be removed  by the  BRST procedure as we will see later.
Hence, the one-loop determinant associated to the fields 
$(\sigma_{Jm}^{\alpha (s,t)},  X_{Jm0}^{\alpha (s,t)})$ is $\det^{-1/2} \Delta^2_{\sigma, X^0}$ 
with
\bea
\det \Delta^2_{\sigma, X^0}= \prod_{J, m, \alpha}  \left( \frac{2}{l}\right)^2 \left( J(J+1) +(\alpha\cdot\sigma )^2 \right).
\eea 
However, this factor is exactly canceled to the factors coming from the one-loop determinant of ghosts and 
gauge fixing delta functions in BRST procedure (See \ref{gaugefixing}). 
Therefore, the one-loop determinants of the transverse parts $X_{\rho =\pm 1}$ give 
the bosonic part of the one-loop determinant of 
the Yang-Mills action $\det^{-1/2}{\Delta_X^2}|_{\rho =\pm 1}  $ with 
\begin{eqnarray}\label{detgauge}
&&\det{\Delta_X^2}|_{\rho =+1}
= \prod_{s,t} \prod_{\alpha\in\Delta} \prod_{J=|j_s -j_t|}^{j_s +j_t} \prod_{m=-(J+1)}^{J+1}
    \left\{ \left( \frac{2}{l}\right)^2 (J+1)^2 +(\alpha\cdot\sigma )^2    \right\}, \NN \\
&&\det{\Delta_X^2}|_{\rho =-1}
=\prod_{s,t} \prod_{\alpha\in\Delta} \prod_{J=|j_s -j_t|-1}^{j_s +j_t-1} \prod_{m=-J}^{J}
    \left\{ \left( \frac{2}{l}\right)^2 (J+1)^2 +(\alpha\cdot\sigma )^2    \right\}. \NN \\
\end{eqnarray}
 
Next, we calculate the contribution from the fermionic fields. 
The Lagrangian of the gaugino at quadratic order is expanded by the spinor fuzzy sphere harmonics as
\begin{eqnarray}\label{YM}
&&   -\frac{1}{l} \bar{\lambda}\gamma^i [\ L_i ,\lambda\ ] 
       +\frac{i}{2}\bar{\lambda} [\ \sigma_0 ,\lambda\ ] -\frac{1}{l}\bar{\lambda}\lambda \NN \\
&=& \sum_{\kappa ,\tilde{U},m}\left\{ -\frac{1}{l} \kappa \left( J+\frac{3}{4}\right) +i \alpha \cdot \sigma  \right\} 
        \Bigl[\  \bar{\lambda}_{Jm \kappa }^{(s,t) \alpha \dag} \lambda_{Jm\kappa}^{\alpha (s,t)}   \ \Bigr].
\end{eqnarray}
By integrating out $\bar{\lambda}_{Jm\kappa}^{(s,t)\dag}, \lambda_{Jm\kappa}^{(s,t)}$, 
one can obtain the fermionic part of the one-loop determinant
$\Delta_\lambda |_{\kappa =\pm 1}$ as
\begin{eqnarray}\label{detgaugino}
\det{\Delta_\lambda |_{\kappa =+1}}
&=& \prod_{s,t} \prod_{\alpha\in\Delta} \prod_{J=|j_s -j_t|}^{j_s +j_t} \prod_{m=-(J-1/2)}^{J+1/2}
    \Biggl[ \frac{2}{l}\left( J+1 \right) -i(\alpha\cdot\sigma )    \Biggr],  \NN \\
\det{\Delta_\lambda |_{\kappa =-1}}
&=& \prod_{s,t} \prod_{\alpha\in\Delta} \prod_{J=|j_s -j_t|-1/2}^{j_s +j_t-1/2} \prod_{m=-J}^{J}
    \Biggl[ \frac{2}{l}\left( -J-\frac{1}{2} \right) -i(\alpha\cdot\sigma )    \Biggr]. \NN \\
\end{eqnarray}

%
\subsection{Matter sector}

\subsubsection*{Localized configuration}
Since the reduced matter action $S_{\rm kin}^r$ itself is rewritten as 
\[
\bar{\epsilon}\epsilon S_{kin}^r
= \delta_{\bar{\epsilon}} \delta_\epsilon 
    \Tr_{M_2} \Bigl[ \bar{\psi}\psi -2i\bar{\phi}(\sigma_{A} -\sigma_{B} )\phi +\frac{2(q-1)}{l}\bar{\phi}\phi \Bigr],
\]
we can choose the deformation term $Q \cdot V$ as the reduced action $S_{kin}^r$. 
From (\ref{redmatter}), the localized configurations are trivial:
\begin{equation}
\phi =\bar{\phi} =\psi =\bar{\psi} =F =\bar{F}=0.
\end{equation}

\subsubsection*{1-loop determinant}
Here we calculate the one-loop determinant of the $U(M_1) \times U(M_2)$ bi-fundamental matter.
We rescale  all the matter component fields as $\Phi \to \frac{1}{\sqrt{t}} \Phi$
and substitute the localization configuration into (\ref{redmatter}).
Then, in the large $t$ limit, only the quadratic part with respect to the matter component survive as follows 
\begin{eqnarray}
\left. \mathcal{L}_{\rm kin}^r \right|_{\rm Gauss}
&=& \Tr_{M_2} \Biggl[
     \left( \frac{2}{l} \right)^2 ( L_i^X \phi -\phi L_i^Y )^\dag  ( L_i^X {\phi} -\phi L_i^Y )
     +\bar{\phi}( \bar{\sigma}^A -\bar{\sigma}^B )^2 \phi \nonumber \\
&&    +\frac{i(2q-2)}{l} \bar{\phi}( \bar{\sigma}^A -\bar{\sigma}^B ) \phi +\frac{q(2-q)}{l^2}\bar{\phi}\phi   \nonumber \\
&&  +\frac{2}{l} \bar{\psi}\gamma^i (L_i^X \psi -\psi L_i^Y ) 
    +i\bar{\psi}(\bar{\sigma}^A -\bar{\sigma}^B )  \psi -\frac{q-2}{l}\bar{\psi}\psi 
    +\bar{F} F \Biggr]. \NN \\
\end{eqnarray}
Here $L^X$ and $L^Y$ are the classical field configuration of reduced gauge fields $X^i$ and $Y^i$, respectively;
\bea
L^X= \bigoplus_{s=1}^{\nu} L_i^{(n_s)} \otimes \mathbf{1}_{N_1}, \quad
L^Y= \bigoplus_{s=1}^{\nu} L_i^{(n_s)} \otimes \mathbf{1}_{N_2}.
\eea
$\bar{\sigma}^A$ and $\bar{\sigma}^B$ are 
$N_1 \times N_1$ and $N_2 \times N_2$ classical scalar field configurations in the vector multiplets.

Expanding $\phi$ and $\psi$ in terms of the fuzzy sphere harmonics as
\bea
&&\phi=\sum_{s, t}        \sum_{J=|j_s -j_t|}^{j_s +j_t} \sum_{m=-J}^J  
                   \hat{Y}_{Jm(j_s j_t)} \otimes  \phi_{Jm}^{(s,t)},  \nonumber \\           
&&\psi_{\beta}  
              = \sum_{s, t} \sum_{\kappa =\pm} \sum_{\tilde{U}=|j_s -j_t|}^{j_s +j_t} \sum_{m=-U}^U 
                 \hat{Y}_{Jm(j_s j_t)\alpha}^{\kappa} \otimes \psi_{Jm\kappa}^{(s,t)},
\eea
and integrating out $ \phi_{Jm}^{(s,t)}$, $\psi_{Jm\kappa}^{(s,t)}$, 
the 1-loop determinants of each field are evaluated in the similar manner to the vector multiplet as
\begin{eqnarray}\label{detmat}
\det{\Delta_{\phi}^2}
&=& \prod_{s,t} \prod_{a=1}^{N_1} \prod_{b=1}^{N_2} \prod_{J=|j_s -j_t|}^{j_s +j_t} \prod_{m=-J}^{J} \NN \\
&&  \Biggl[ \left( \frac{2}{l}\right)^2 J(J+1) +(\sigma_{a} -\tilde{\sigma}_{b})^2
           +\frac{i(2q-2)}{l}(\sigma_{a} -\tilde{\sigma}_{b}) +\frac{q(2-q)}{l^2}
    \Biggr], \nonumber \\
\det{\Delta_\psi |_{\kappa =+1}}
&=& \prod_{s,t} \prod_{a=1}^{N_1} \prod_{b=1}^{N_2} \prod_{J=|j_s -j_t|}^{j_s +j_t} \prod_{m=-(J-1/2)}^{J+1/2}
    \Biggl[ \frac{2}{l} J +i(\sigma_{a} -\tilde{\sigma}_{b} ) -\frac{q-2}{l}  \Biggr], \nonumber \\
\det{\Delta_\psi |_{\kappa =-1}}
&=& \prod_{s,t} \prod_{a=1}^{N_1} \prod_{b=1}^{N_2} \prod_{J=|j_s -j_t|-1/2}^{j_s +j_t-1/2} \prod_{m=-J}^{J}
   \Biggl[ \frac{2}{l}\left( -J-\frac{3}{2} \right) +i(\sigma_{a} -\tilde{\sigma}_{b} ) -\frac{q-2}{l}    \Biggr] , \NN \\ 
\end{eqnarray}
where we take the diagonal gauge $\sigma_0^A =\mathrm{diag}(\sigma_1,\cdots, \sigma_{N_1})$
and $\sigma_0^B =\mathrm{diag}(\tilde{\sigma}_1,\cdots, \tilde{\sigma}_{N_2})$.

The one-loop determinant of the bi-fundamental matter multiplet is given by
\bea\label{detmatter}
\frac{\det{\Delta_\psi |_{\kappa =+1}} \det{\Delta_\psi |_{\kappa =-1}} }{\det{\Delta_{\phi}^2}}. 
\eea

\section{Planar Large N reduction}
\label{sec:planar}
In this section, we show the large N reduction for the free energy and the BPS Wilson loops in the planar limit.

\subsection{Free energy}
Here we consider the large N equivalence for the free energy.
We consider the $G=\prod_{I=1}^{A} U(M_I)_{g_I}$ reduced quiver CSM theories 
with any number of the bi-fundamental and adjoint matters.  
For a later convenience, we introduce $g_I:=g_{CS,r I} V^{1/2}_{S^3}$,
where $g_{CS,r I}$ is the coupling constant of the $U(M_I)$ reduced CS theory.
Let $m_{I J}$ be the number of the chiral multiplets with the bi-fundamental representation 
under $U(M_I) \times U(M_J)$ for $I \neq J$ or with adjoint representation for $I= J$.
We take the following limit 
\bea\label{largen}
\ N_I \to \infty, \ \nu \to \infty, \frac{n}{\nu}\ \RA\ \infty \ 
\mathrm{with} \ \lambda_I =\frac{g_I^2 N_I}{n} = \frac{4\pi N_I}{ k_I}={\mathrm{fixed}},\ 
\frac{N_I}{N_J}={\mathrm{fixed}},
\eea
which corresponds to the 't Hooft limit in the original theory.
If we define the free energy in the original theory and the reduced model
as $F^{3d} = \log Z^{3d}$ and $F^{r} = \log Z^r$, respectively,
the statement of the large N reduction on $S^3$ is
\bea\label{equivalence}
\left. F^{3d}\right|_{\rm planar}  = \left. \frac{F^{r}}{\nu } \right|_{\rm planar}.
\eea
For the $U(N)$ pure CS theory, this relation is shown 
by using the Feynman diagram technique \cite{Ishiki:2009vr,Ishiki:2010pe}.
In the  limit $\nu\ \RA\infty ,\ n/\nu \RA \infty$,
combining (\ref{detgauge}) and (\ref{detgaugino}), 
the one-loop determinant for the vector multiplet is given by\footnote{
For the detail, see \ref{sec:det}.}  
\bea\label{rymdeter}
&& (\alpha\cdot\sigma )^2 \frac{ \det \Delta_{\lambda} |_{\kappa =+1} \cdot \det \Delta_{\lambda} |_{\kappa =-1} } 
     { \sqrt{\det \Delta^2_{ X}|_{\rho =+1}} \cdot \sqrt{\det \Delta^2_{ X}|_{\rho =-1}}   } \NN \\
&\simeq& \prod_{\alpha\in\Delta_+}  \left\{ (\alpha\cdot\sigma )^2 \right\}^{2\nu}
   \prod_{n=1}^\infty  \left\{ \left( \frac{n}{l}\right)^2 +(\alpha\cdot\sigma )^2     \right\}^{2\nu} \non
&=&  \prod_{1\le a <b \le N_I}
  2^{2\nu}  \sinh^{2\nu} (\pi l (\sigma^I_a -\sigma^I_b ) ).
\label{ymldet}
\eea
From (\ref{detmat}), the contribution from the the bi-fundamental matter is 
\bea\label{mattldet}
\frac{\det{\Delta_\psi |_{\kappa =+1}} \det{\Delta_\psi |_{\kappa =-1}} }{\det{\Delta_{\phi}^2}}
&\simeq& \prod_{a ,b}  \prod_{n=1}^\infty \Biggl[
     \frac{\frac{n+1}{l} -\frac{q}{l} +i(\sigma^I_{a} -{\sigma}^J_{b}) }
          {\frac{n-1}{l} +\frac{q}{l} -i(\sigma^I_{a} -{\sigma}^J_{b}) }
      \Biggr]^{n\nu} \nonumber \\
&=& \prod_{a=1}^{N_I} \prod^{N_J}_{b=1} s_{b=1}^\nu (i-iq -l(\sigma^I_{a} -{\sigma}^J_{b} ) ),
\end{eqnarray}
where $s_b(x)$ is the double sine function defined by
\bea
s_b(x):=\prod_{m,n=0}^{\infty} \frac{mb+n b^{-1}+\frac{Q}{2}-ix}{mb+n b^{-1}+\frac{Q}{2}+ix}, \quad \mathrm{with} \quad   Q:=b+\frac{1}{b}.
\eea
Here we note that 
the one-loop determinants of the original $U(N_I)$ $\mathcal{N}=2$ super Yang-Mills theory 
and the $U(N_I) \times U(N_J)$ bi-fundamental chiral multiplet are 
$ \prod_{1\le a <b \le N_I} 2^{2} \sinh^{2} (\pi l (\sigma^I_a -\sigma^I_b ) ) $ 
and $ \prod_{a=1}^{N_I} \prod^{N_J}_{b=1} s_{1} (i-iq -l(\sigma^I_{a} -{\sigma}^I_{b} ) )$.   
Therefore, the one-loop determinants of the reduced model are the $\nu$-th  power of the original one.  
From (\ref{ymldet}) and  (\ref{mattldet}),
the partition function of the $G=\prod_{I=1}^{A} U(M_I)_{g_I}$ reduced supersymmetric quiver Chern-Simons matter theory is
\bea\label{reducedpartition}
Z^{r}&=&\int \prod_{I=1}^{A} \prod_{a=1}^{N_I}   
d \sigma^I_a  \Delta^{ \nu}(\sigma) 
\exp  \left( \nu \sum_{I=1}^{A} \sum_{a=1}^{N_I} \frac{i N_I}{\lambda_I }  \sigma^{I 2}_a \right),
\eea
where the $\Delta(\sigma)$ is same as 
the one-loop determinant of the $G=\prod_{I=1}^{A} U(N_I)_{k_I}$ three dimensional quiver supersymmetric CSM given by
\bea
\Delta(\sigma)=\prod_{I} \prod_{ a <b  } 2^{2}
\sinh^2 (\pi l (\sigma^{I}_a-\sigma^{I}_{b} ) ) 
\prod_{I, J} \prod_{a,b} s^{m_{I J}}_{1} (i-iq -l(\sigma^{I}_{a} -{\sigma}^{J}_{b} ) ).
\eea 
In order to discuss the correspondence for the planar free energy,
we write down the saddle point equation of the reduced model as 
\bea\label{saddle}
0=\frac{\partial S^{r}_{\mathrm{eff}}}{\partial \sigma^{I}_a}(\sigma^*)
=\nu \Bigl(  \partial_{\sig_I} \log \Delta (\sigma^*)  +  \frac{2i N_I }{\lambda_I  }  \sigma^{I *}_{a} \Bigr) ,
\eea
where $S^{r}_{\mathrm{eff}}$ is the effective action given by
\bea
S^{r}_{\mathrm{eff}} 
=\nu \left(  \log  \Delta(\sigma)+ \sum_{I=1}^{A} \sum_{a=1}^{N_I} \frac{i N_I}{\lambda_I }  \sigma^{I 2}_a \right).
\eea
The planar free energy is dominated by the solutions of the saddle point equation
\bea
\left. F^r \right|_{\rm planar}  =S^r_{\mathrm{eff}} (\sigma^*).
\eea
In order to write down the planar free energy more explicitly, 
we introduce the  eigenvalue density $\rho^I (x)$ for $\sigma^I=\mathrm{diag}(\sigma^I_1, \cdots, \sigma^I_N)$ 
\bea
\rho_I (x):=\frac{1}{N_I} \sum_{a=1}^{N_I} \langle \delta (x-\sigma^{I *}_a) \rangle.
\eea
Here  $\sigma^{I *}_a$ are the solutions of (\ref{saddle}),
In the large $N_I$ limit, $\rho_I (x)$ approaches to the continuous distribution with the normalization condition
\bea
\int_{\mathcal{C}_I} d x \rho_I (x) =1.
\eea
Since $S_{\rm eff}^r$ is proportional to the original one,
the saddle point equations for the reduced model (\ref{saddle}) are same for  the $U(N_I)$ quiver CSM matrix model.
Therefore the resolvents and the eigenvalue densities  of the two theories are identical with each other.
In the $N_I \to \infty$ limit,  the summation is goes to integrand
\bea
\lim_{N_I \to \infty} \frac{1}{N_I} \sum_{a=1}^{N_I} f(\sigma^I_a)=\int_{\mathcal{C}_I} dx f(x).
\eea
Thus, the planar free energy is described by the effective action $S^{r}_{\mathrm{eff}}$ 
with respect to the saddle point (\ref{saddle}) as
\bea
\left. F^{r} \right|_{\rm planar} 
&=&\nu \Biggl[  \sum_{I}  \frac{i N_I^2 }{\lambda_I }  \mathrm{p.v} \int_{\mathcal{C}_I } dx  x^{ 2} \rho_I (x)  \NN \\
&& +\sum_{I } N_I^2 \int_{\mathcal{C}_I \times \mathcal{C}_I} dx dx'  
   \log 2^{2} \sinh^2 ( \pi l (x-x' ) ) \rho_I(x) \rho_I(x')   \nonumber \\
&& +\sum_{I, J}  N_I N_J 
   \int_{\mathcal{C}_I \times \mathcal{C}_J } dx  dx' 
      \log s^{m_{IJ}}_{b=1} (i-iq_m -l(x -x' ) ) \rho_I(x) \rho_J(x') \Biggr] \NN \\
&=& \left. \nu F^{\rm 3d} \right|_{\rm planar},
\eea
which is nothing but (\ref{equivalence}).
Thus we show the large N equivalence for the planar free energy.

\subsection{BPS Wilson loops}
Here we consider the large N equivalence for the BPS Wilson loops.

\subsubsection*{BPS Wilson loop preserving 2 supercharges}
We consider the following type of the Wilson loop 
\bea\label{wilson}
{W}_R(C)
:=\frac{1}{\mathrm{dim} R} \tr_{R} P 
   \exp \Bigl( \oint_C  d \tau \bigl( i X_i (\Omega_3 ) e^{i}_{\mu}(\tau) dx^{\mu} +\sigma |\dot{x}(\tau)| \bigr)  \Bigr).
\eea
Here $R$ denotes the representation under the $U(N)$ gauge group and $\tr_{R}$ is taken over the representation $R$. 
When the integration contour is the great circle of $S^3$, 
${W}_R(C)$ preserves two supercharges \cite{Drukker:2008zx,Chen:2008bp,Rey:2008bh}.
For example, if we embed ${W}_R(C)$ into the ABJ(M) model \cite{Aharony:2008ug,Aharony:2008gk} 
and integrate out auxiliary scalar $\sigma$, 
we can regard the Wilson loop (\ref{wilson}) as $1/6$-BPS Wilson loop \cite{Benna:2008zy}. 
It has been conjectured that 
the Wilson loop with the different representation $R$ corresponds to a different quantity on the gravity side.
For instance, the gauge/gravity duality states that
the Wilson loops with the fundamental, symmetric and anti-symmetric representation are dual to
the string, D2-brane and D6-brane worldvolume, respectively \cite{Drukker:2008zx}.

If the contour is the great circle, then we can choose angular velocity $ |\dot{x}(\tau)|=1$ without loss of generality. 
The expectation value of the Wilson loop is defined by
\bea
\label{wilson16}
\langle {W}_R(C) \rangle := \frac{1}{Z^{3d}} \int \mathcal{D} \mu  {W}_R(C) \exp (-S^{3d}).
\eea
Here $\mathcal{D} \mu$ is the integration measure of the supersymmetric CSM theory.
Correspondingly, we consider the following operator in the reduced model 
\bea
\hat{W}_{\tilde{R}}(C)
:=\frac{1}{  \mathrm{dim} \tilde{R}} 
  \Tr_{\tilde{R}} P 
    \exp \Bigl( \oint_C  d \tau \bigl( i X_i e^{i}_{\mu}(\tau) dx^{\mu} +\sigma |\dot{x}(\tau)| \bigr)    \Bigr).
\eea
Here the representation $\tilde{R}$ is defined by $\tilde{R}:=\bigoplus_{I=1}^{\nu} \mathbf{1}_{n_I}  \otimes R$. 
The dimension of $\tilde{R}$ is 
$\mathrm{dim} \tilde{R}=n \nu \dim{R}$.
In \cite{Ishii:2007sy}, it has been shown that 
this operator corresponds to the angular average of the original BPS Wilson loop operator in the 't Hooft limit. 
The operator at localization point can be written as
\bea
 \hat{W}_{\tilde{R}}(C)  &=&\frac{1}{  \mathrm{dim} \tilde{R}} \Tr_{R} 
 \exp \Bigl(   i \bigoplus_{s=1}^{\nu} L_i^{(n_s)} \otimes \mathbf{1}_{N}  \oint_C  d \tau e^{i}_{\mu}(\tau) dx^{\mu} 
     +2 \pi l \bigoplus_{I=1}^{\nu} \mathbf{1}_{n_I} \otimes \sigma_0    \Bigr) \nonumber \\
&=&\frac{1}{  \mathrm{dim} \tilde{R}} \Tr_{\tilde{R}} 
 \exp \Bigl(  2 \pi l \bigoplus_{I=1}^{\nu} \mathbf{1}_{n_I} \otimes\sigma_0   \Bigr) \nonumber \\
 &=&\frac{1}{ \mathrm{dim} R} \tr_{R} 
 \exp \Bigl(  2 \pi l \sigma_0      \Bigr) .
\eea
Note that this is same as the Wilson loop in the original theory at the localization configuration \cite{Kapustin:2009kz}. 
The expectation value of the Wilson loop is defined by
\bea
\langle  \hat{W}_{\tilde{R}}(C) \rangle := \frac{1}{Z^{r}} \int d X_i d \lambda \cdots  \hat{W}_{\tilde{R}}(C) \exp (-S^r)
\eea
From the localization argument, the expectation value of the BPS Wilson loop is given by
\bea
\langle  \hat{W}_{\tilde{R}}(C) \rangle 
= \frac{1}{ \mathrm{dim} R Z^{r}}  
  \int \prod_{I=1}^{A} \prod_{a=1}^{N_I}  d \sigma^I_a  \Delta^{ \nu}(\sigma) \tr_{R} 
  \exp \Bigl(  2 \pi l \sigma_{0}      \Bigr) 
  \exp \left( \nu \sum_{I=1}^{A} \sum_{a=1}^{N_I} \frac{i N_I}{\lambda_I }  \sigma^{I 2}_a \right).
\eea
In the large $N_I$ limit, 
the expectation value of the Wilson loop can be also evaluated by the solutions of the saddle points equation
(\ref{saddle})\footnote{We assume the rank of the representation 
is small and the saddle points are unaffected by the insertion of Wilson loop operator. }
and becomes
\bea
\langle  \hat{W}_{\tilde{R}}(C) \rangle 
= \frac{1}{ \mathrm{dim} R }  \int dx\  \rho(x)   \tr_{R}  \exp \Bigl(  2 \pi l x      \Bigr)
=\langle {W}_R(C) \rangle .
\eea
Thus we show the large N correspondence for the Wilson loop (\ref{wilson16}) in the planar limit.

\subsubsection*{1/2-BPS Wilson loop in the ABJ(M) theory}
In \cite{Drukker:2009hy}, 
the authors construct the 1/2-BPS Wilson loop in the ABJ(M) theory, whose form is given by
\bea\label{wilson12}
{SW}_R(C)
:=\frac{1}{\mathrm{dim} R} \tr_{R} P 
   \exp \Bigl( \oint_C  d \tau L  \Bigr),
\eea
where $L$ is 
\begin{\eq}
L := 
\begin{pmatrix}
iX_i (\Omega_3 ) e^{i}_{\mu}(\tau) \dot{x}^{\mu} +\sigma_A (\Omega_3 )|\dot{x}(\tau)|  
                            & i\sqrt{\frac{2\pi}{k}} |\dot{x}(\tau)| \eta^\alpha_I (\tau )\bar{\psi}_\alpha^I (\Omega_3 )\cr
i\sqrt{\frac{2\pi}{k}} |\dot{x}(\tau )| \psi_I^\alpha (\Omega_3 ) \bar{\eta}_\alpha^I (\tau ) & 
                                       iY_i (\Omega_3 ) e^{i}_{\mu}(\tau) \dot{x}^{\mu} +\sigma_B (\Omega_3 )|\dot{x}(\tau)|  
\end{pmatrix}.
\end{\eq}
Although $\eta^\alpha_I (\tau )$ and $\bar{\eta}^\alpha_I (\tau )$ are the parameters 
determined by requiring supersymmetry,
these are irrelevant on the localized configuration.
Correspondingly, we consider the operator in the reduced model:
\bea
{S\hat{W}}_{\tilde{R}}(C)
:=\frac{1}{\mathrm{dim} \tilde{R}} \tr_{\tilde{R}} P 
   \exp \Bigl( \oint_C  d \tau \hat{L}  \Bigr),
\eea
where $\hat{L}$ is given by
\begin{\eq}
\hat{L} := 
\begin{pmatrix}
iX_i  e^{i}_{\mu}(\tau) \dot{x}^{\mu} +\sigma_A |\dot{x}(\tau)|  
                                   & i\sqrt{\frac{2\pi}{k}} |\dot{x}(\tau)| \eta^\alpha_I (\tau )\bar{\psi}_\alpha^I \cr
i\sqrt{\frac{2\pi}{k}} |\dot{x}(\tau )| \psi_I^\alpha \bar{\eta}_\alpha^I (\tau )  & 
                                          iY_i  e^{i}_{\mu}(\tau) \dot{x}^{\mu} +\sigma_B |\dot{x}(\tau)| 
\end{pmatrix}.
\end{\eq}

This operator at the localization point can be written as
\bea
 \hat{W}_{\tilde{R}}(C)  
&=&\frac{1}{ \mathrm{dim} R} {\rm Str}_{R} 
   \begin{pmatrix}
   e^{2\pi l\sigma_0^A}  & 0 \cr
      0     & -e^{2\pi l\sigma_0^B}
    \end{pmatrix},
\eea
which is same as the one in the original theory at the localized configuration \cite{Drukker:2009hy}.
From the localization argument, the expectation value of the BPS Wilson loop 
\begin{\eqa}
\langle  S\hat{W}_{\tilde{R}}(C) \rangle 
&=& \frac{1}{ \mathrm{dim} R Z^{r}}  
 \int \prod_{I=1}^{2} \prod_{a=1}^{N_I}  d \sigma^I_a  \Delta^{ \nu}(\sigma)   
{\rm Str}_{R} 
   \begin{pmatrix}
   e^{2\pi l\sigma_0^A}  & 0 \cr
      0     & -e^{2\pi l\sigma_0^B}
    \end{pmatrix} \NN \\
&& \times \exp  \left( \nu \sum_{I=1}^{2} \sum_{a=1}^{N_I} \frac{i N_I}{\lambda_I }  \sigma^{I 2}_a \right)
\end{\eqa}
In the planar limit, 
the expectation value of the Wilson loop can be also evaluated by the solutions of the saddle points equation,
which is same as the original one.
Therefore, we conclude that this is same for the expectation value of the Wilson loop in three dimensions.

\section{M-theoretic Large N reduction}
\label{sec:M-theory}
So far, we have considered the large N reduction on $S^3$ 
for the planar limit: $N_I \rightarrow \infty ,\frac{N_I}{k_I}={\rm fixed}$.
In this limit, many supersymmetric quiver Chern-Simons theories
have been conjectured to be dual to the type-IIA string theory on certain backgrounds.
However, the gauge/gravity duality suggests that
these theories are also dual to M-theory on certain backgrounds 
for the M-theory limit: $N_I \rightarrow \infty ,k_I ={\rm fixed}$.
Let us consider the large N reduction on $S^3$ in the M-theory limit.
This is highly nontrivial in the following reason.
First of all, we do not well understand  general properties of the field theory in the M-theory limit  
although there are recently a few developments \cite{Herzog:2010hf,Marino:2011eh}.
Secondly we emphasize that 
any perturbative argument is unavailable for justifying the large N reduction in the M-theory limit.
Finally, it is nontrivial in this limit whether an usual large N factorization  as for the planar limit occurs or not.
Therefore, we expect that an usual argument by Schwinger-Dyson equation \cite{Eguchi:1982nm} is not also useful.
Now we can perform non-perturbative argument thanks to the localization method.
Let us consider the following limit in the reduced model: 
\begin{\eq}
N\ \RA\ \infty ,\quad \nu\ \RA\ \infty ,\quad \frac{n}{\nu}\ \RA\ \infty \quad 
{\rm with}\quad \frac{g_I^2}{n} =\frac{4\pi}{k_I}={\rm fixed},
\label{red_Mlimit}
\end{\eq}
which is the counterpart of the M-theory limit in the original theory. 
In this limit, the partition function of the reduced model becomes
\bea
Z^{r}
&=&\int \prod_{I=1}^{A} \prod_{a=1}^{N_I} d \sigma^I_a  
\Delta^{ \nu}(\sigma) \exp  \left( \nu \sum_{I=1}^{A} \sum_{a=1}^{N_I} \frac{i}{g_I^2}  \sigma^{I 2}_a \right)
\eea
which is formally same as the planar partition function (\ref{reducedpartition}) 
if we replace $g_I^2 =\frac{\lambda_I}{N_I}$.
As in \cite{Herzog:2010hf}, we assume that a saddle point for the integral exists in the limit (\ref{red_Mlimit}).
The saddle point equation is 
\bea
\label{saddle_M}
\partial_{\sig_I} \log \Delta (\sigma^*)  +  \frac{2i}{g_I^2}  \sigma^{I *}_{a}  =0,
\eea
which is same as the one of the original theory\footnote{
While a standard analysis in the planar limit assumes that edges of eigenvalue distributions is $O(1)$,
we must consider that edges of eigenvalue distributions in the M-theory limit depends on $N_I$ 
in order to obtain a nontrivial solution. See the \cite{Herzog:2010hf} for detail.
} in the M-theory limit \cite{Herzog:2010hf} 
if we substitute $\frac{g_I^2}{n} =\frac{4\pi}{k_I}$.
Therefore, we conclude that the relation of the free energy between the reduced model and the original model is
\begin{\eq}
\frac{F^r}{\nu}=F^{3d},
\end{\eq} 
also for the M-theory limit.
Similarly, the correspondence also holds for the BPS Wilson loops (\ref{wilson}) and (\ref{wilson12}) in the limit
since the saddle points are same with each other.

\section{Conclusion}
\label{sec:conclusion}
In this paper,   
we show the large N reduction on $S^3$ in the BPS sector
for the general $\mathcal{N}\geq 2$ supersymmetric quiver Chern-Simons theories by using the localization method.
In particular, we calculate the free energy, expectation values of 1/6-BPS Wilson loops and 1/2-BPS Wilson loops.
Remarkably, it turns out that the large N reduction holds even for the M-theory limit.
Although we have shown that formally, we ask ``What does this mean physically ?''.
We are suspicious to that this is related to a smooth connection between the planar and M-theory limit,
which has been recently observed 
for the free energy \cite{Drukker:2010nc,Herzog:2010hf,Drukker:2011zy,Fuji:2011km,Marino:2011eh,Hanada:2012si} 
and BPS Wilson loops \cite{Marino:2009jd} in the ABJM theory.
These results imply that 
the free energy and BPS Wilson loops in the strong 't Hooft coupling regime after taking the 't Hooft limit 
are same as the ones in the M-theory limit if we simply replace $\lambda =N/k$.
Therefore, we expect that 
such a smooth connection between the 't Hooft and M-theory limit is 
one of sufficient conditions for the large N reduction in the M-theory limit.
 
We remark on an important point 
that the reasons why our localization in the reduced model works well are following.
When we performed dimensional reduction of the theories on $S^3$, 
flat directions of the reduced gauge fields in the $Q$-exact term disappear 
and emerge the non-trivial fuzzy sphere solution at the localization points.  
This is quite different from the localization of the reduced models on flat space 
which suffer from the divergence coming from the flat directions. 
We expect that our localization method 
is useful to show the large N reduction in other theories defined on $S^3$; 
In \cite{Ishii:2008tm}, the large N equivalence between the $\mathcal{N}=4$ super Yang-Mills theory on $\R \times S^3$ 
and the plane wave (BMN) matrix model \cite{Berenstein:2002jq} on $\R$ has been proposed. 
The localization of this system will be studied in our future work.

Applications to the same class of theory on different space would be also interested.
For instance, the theories on $S^1 \times S^2$ and squashed $S^3$ 
are studied in \cite{Kim:2009wb,Imamura:2011uj,Hama:2011ea,Imamura:2011wg}.
In particular, since the squashed $S^3$ is neither compact semi-simple group manifold nor its coset space
whose reduced model is constructed in \cite{Kawai:2009vb,Kawai:2010sf},
it may give some insights to emergent geometry.

\subsubsection*{Note added}
When our paper was ready for submission to the arXiv,
there appeared a paper \cite{Asano} which has overlap with ours.

\vspace{1.3cm}
\centerline{\bf Acknowledgements}
We would like to thank Kazuo Hosomichi for a enlightening lecture 
on the localization method in supersymmetric gauge theories 
at Chubu Summer School 2011. 
We are grateful to Masanori Hanada, Hikaru Kawai, So Matsuura, Jun Nishimura, Kazutoshi Ohta, Norisuke Sakai,
Fumihiko Sugino and Asato Tsuchiya for helpful discussions.
M.H is grateful to the hospitality 
of the Kavli Institute for Theoretical Physics in UCSB,
during the KITP program 
``Novel Numerical Methods for Strongly Coupled Quantum Field Theory
and Quantum Gravity''. 
The work of M.\ H.\ is supported
by Japan Society for the Promotion of Science (JSPS).

\appendix
\section{$S^3$ as $SU(2)$ group manifold}
In this appendix, we summarize properties of $S^3$.
Let us start with the parametrization of the unit $S^3$ by an element $g$ of the Lie group $SU(2)$:
\begin{\eqa}
g
&=& e^{i\alpha\gamma_3}e^{i\theta\gamma_2}e^{i\beta\gamma_3} \\
&=& \begin{pmatrix}
     e^{ i(\alpha +\beta )} \cos{\theta}  & -e^{ i(\alpha -\beta )} \sin{\theta} \cr
    -e^{-i(\alpha -\beta )} \sin{\theta}  &  e^{-i(\alpha +\beta )} \cos{\theta} 
    \end{pmatrix},  
\end{\eqa}
where $\gamma^i \ (i=1,2,3) $ are Pauli matrices and
$0\leq\theta\leq\frac{\pi}{2},0\leq\alpha\leq\pi ,0\leq\beta\leq 2\pi$.
The left invariant 1-form $\mu$ is defined by
\begin{\eq}
g^{-1}dg = i\mu^i \gamma^i ,
\end{\eq}
which satisfies the Maurer-Cartan equation $d\mu^i = \epsilon^{ijk}\mu^j \wedge \mu^k$.
The left invariant 1-form $\mu$ gives the metric of $S^3$ with the radius $l$ as
\begin{\eqa}
ds^2 
&=& \frac{1}{2}l^2\tr \left( dg dg^{-1}\right) =l^2 \mu^i \mu^i \nonumber \\
&=& l^2 \Bigl[ d\theta^2 +d\alpha^2 +(d\beta +\cos{2\theta}d\alpha )^2 \Bigr] .
\end{\eqa} 
Defining the dreibein in the left-invariant frame as $e^i =l\mu^i$,
each component of $e$ is explicitly given by
\begin{\eqa}
e^1 &=& l\left( -\sin{2\beta}d\theta +\cos{2\beta}\sin{2\theta}d\alpha \right), \nonumber \\ 
e^2 &=& l\left(  \cos{2\beta}d\theta +\sin{2\beta}\sin{2\theta}d\alpha \right), \nonumber \\ 
e^3 &=& l\left(  d\beta              +\cos{2\theta}            d\alpha \right), 
\end{\eqa}
where $e^i$ satisfies $de^i = \frac{1}{l}\epsilon^{ijk}e^j \wedge e^k$ from the Maurer-Cartan equation.
The spin connection in this frame is 
\begin{\eq}
\omega^{ij} = \frac{1}{l}\epsilon^{ijk}e^k .
\end{\eq}
The Killing vector $J_i$ dual to $e^i$ is
\begin{\eq}
J_i = \frac{l}{2i}e_i^a \partial_a
\end{\eq}
where $a=\theta ,\alpha ,\beta$ and $e_i^a$ is the inverse of $e^i_a$.
The explicit form of each $J_i$ is
\begin{\eqa}
J_1 &=& \frac{1}{2i} 
         \left( -\sin{2\beta}\partial_\theta 
                 +\frac{\cos{2\beta}}{\sin{2\theta}}\partial_\alpha -\cot{2\theta}\cos{2\beta}\partial_\beta  \right)  \nonumber \\
J_2 &=& \frac{1}{2i} 
         \left(  \cos{2\beta}\partial_\theta 
                 +\frac{\sin{2\beta}}{\sin{2\theta}}\partial_\alpha -\sin{2\theta}\cos{2\beta}\partial_\beta  \right)  \nonumber \\
J_3 &=& \frac{1}{2i} \partial_\beta
\label{KillingS3}
\end{\eqa}
and these satisfy $SU(2)$ algebra
\begin{\eq}
[\ J_i , J_j\ ]=i\epsilon_{ijk}J_k   .
\end{\eq}

\section{Fuzzy sphere harmonics}
\label{sec:fuzzy}
In this section we briefly review the definition of the fuzzy sphere harmonics and their basic properties.
Let $L_i \ (i=1,2,3)$, $L^{(n)}_i$ and $| j, m \rangle, (m=-j, -j+1, \cdots,j) $
be the generator of $SU(2)$, the $n=2j+1$ dimensional irreducible representation 
and a basis of the representation space, respectively. 

Then the tensor products $|j, m \rangle \langle j',m'|$ span a basis of $(2j+1) \times (2j'+1)$-dimensional  representation.
$L_i$ acts on  $|j, m \rangle \langle j',m'|$ as 
\bea
L_i \circ |j, m \rangle \langle j',m'|:=L^{(n)}_i |j, m \rangle \langle j',m'|- |j, m \rangle \langle j',m'| L^{(n')}_i,
\eea
where $n=2j+1$ and $n'=2j'+1$.
The scalar fuzzy sphere harmonics is defined by
\bea
\hat{Y}_{Jm(jj^\prime )}:=\sqrt{n} \sum_{r, r'} (-1)^{-j+r'} C^{Jm}_{jr j' -r'} |j, m \rangle \langle j',m'|. 
\eea
Here $C^{Jm}_{jr j' -r'}$ are the Clebsch-Gordan coefficients.
The fuzzy sphere harmonics also spans a basis of the tensor representation. 
The vector and spinor spherical harmonics are defined by
\bea
&&\hat{Y}^{\rho}_{Jm(jj^\prime ) i}:=i^{\rho} V_{i n} C^{Qm}_{\tilde{Q} p 1 n} \hat{Y}_{\tilde{Q} p(jj^\prime )}, \\
&&\hat{Y}^{\kappa}_{Jm(jj^\prime ) \alpha}:= C^{Um}_{\tilde{U} p \frac{1}{2} \alpha} \hat{Y}_{\tilde{Q} p(jj^\prime )}.
\eea
Here $\alpha=1,2$ and  $Q=J+\frac{(1+\rho )\rho}{2},\tilde{Q}=J-\frac{(1-\rho )\rho}{2}, U=J+\frac{1+\kappa}{4}, \tilde{U}=J+\frac{1-\kappa}{4}$
with $\rho=-1, 0, 1$ and $\kappa=\pm1$. $V$ is $3 \times 3$ matrix defined by
\bea
V=
\begin{pmatrix} 
                             -1 & 0 & 1\\
                               -i& 0 & -i  \\ 
                                 0& \sqrt{2} & 0                
                              \end{pmatrix}. 
\eea
The fuzzy sphere harmonics satisfies the following useful formula,
\begin{eqnarray}\label{commu}
&&L_{\pm}\circ \hat{Y}_{Jm(jj^\prime )}        = \sqrt{(J\mp m)(J\pm m+1)} \hat{Y}_{Jm\pm 1(jj^\prime )}, \nonumber \\
&&L_3 \circ \hat{Y}_{Jm(jj^\prime )}           = m \hat{Y}_{Jm(jj^\prime )}, \nonumber \\
&&L_i \circ L_i \circ \hat{Y}_{Jm(jj^\prime )} =J(J+1) \hat{Y}_{Jm(jj^\prime )}, \nonumber \\
&&L_i \circ \hat{Y}_{Jm(jj^\prime ) }= \sqrt{J(J+1)}   \hat{Y}^0_{Jm(jj^\prime ) i}, \nonumber \\
&&L_i \circ \hat{Y}^{\rho}_{Jm(jj^\prime ) i}= \sqrt{J(J+1)} \delta_{\rho 0}  \hat{Y}_{Jm(jj^\prime )}, \nonumber \\
&&i \epsilon_{ikn} L_{k} \circ \hat{Y}^{\rho}_{J m(j j') n}+\hat{Y}^{\rho}_{J m(j j') i}=\rho(J+1) \hat{Y}^{\rho}_{J m(j j') i}, \nonumber \\
&&\Bigl( \gamma^{i}_{\alpha \beta}  L_i \circ +\frac{3}{4} \delta_{\alpha \beta} \Bigr) \hat{Y}^{\kappa}_{Jm(jj^\prime ) \beta}=\kappa(J+\frac{3}{4}) \hat{Y}^{\kappa}_{Jm(jj^\prime ) \alpha}.
\end{eqnarray}
Here $\gamma^{i}, (i=1, 2, 3)$ are the Pauli matrices.
The orthogonal relations of the fuzzy sphere harmonics are   
\begin{eqnarray}
&&\frac{1}{n}\tr\left( \hat{Y}_{J_1 m_1(jj^\prime )}^\dag \hat{Y}_{J_2 m_2(jj^\prime )} \right) = \delta_{J_1 J_2}\delta_{m_1 m_2}, \nonumber \\
&&\frac{1}{n}\tr\left( \hat{Y}^{\rho_1 \dag}_{J_1 m_1(jj^\prime ) i} \hat{Y}^{\rho_2}_{J_2 m_2(jj^\prime ) i} \right) =
\delta_{\rho_1 \rho_2} \delta_{J_1 J_2}\delta_{m_1 m_2}, \nonumber \\
&&\frac{1}{n}\tr\left( \hat{Y}^{\kappa_1 \dag}_{J_1 m_1(jj^\prime ) i} \hat{Y}^{\kappa_2}_{J_2 m_2(jj^\prime ) i} \right) =
\delta_{\kappa_1 \kappa_2} \delta_{J_1 J_2}\delta_{m_1 m_2}.
\end{eqnarray}

\section{Supersymmetry}
\label{sec:SUSY}

In this appendix, 
we summarize the  supersymmetric transformations of 
the $\mathcal{N}=2$ super CSM theories on $S^3$ and their reduced versions. 
These theories  consist of the $\mathcal{N}=2$ vector multiplets and the matter chiral multiplets.
\subsection{Gauge sector}
The $\mathcal{N}=2$ CS action $S_{\rm CS} $ and SYM action $S_{YM}$ on $S^3$ 
are invariant under the supersymmetric transformation \cite{Kapustin:2009kz}:
\begin{eqnarray}
\delta A_a     &=& -\frac{i}{2} ( \bar{\epsilon}\gamma_a \lambda -\bar{\lambda} \gamma_a \epsilon ) , \NN \\ 
\delta \sigma  &=&  \frac{1}{2} ( \bar{\epsilon}\lambda -\bar{\lambda} \epsilon ) , \NN \\ 
\delta \lambda &=& \frac{1}{2} \gamma^{ab}\epsilon F_{ab} -D\epsilon +i\gamma^a \epsilon D_a \sigma +2i\sigma\tilde{\epsilon} ,
                    \NN \\
\delta \bar{\lambda} &=&  \frac{1}{2} \gamma^{ab}\bar{\epsilon}F_{ab} +D\bar{\epsilon}
                         -i\gamma^a \bar{\epsilon} D_a \sigma -i\sigma\tilde{\bar{\epsilon}} ,
                    \NN \\
\delta D       &=& -\frac{i}{2} \bar{\epsilon}\gamma^a D_a \lambda -\frac{i}{2} D_a \bar{\lambda}\gamma^a \epsilon
                   +\frac{i}{2} [ \bar{\epsilon}\lambda +\bar{\lambda}\epsilon ,\sigma ]
                   +\frac{i}{2} ( \tilde{\bar{\epsilon}}\lambda -\bar{\lambda}\tilde{\epsilon} ) , 
\end{eqnarray}
where $\epsilon, \tilde{\epsilon}$ satisfy the killing spinor equation:
\begin{\eq}
\tilde{\epsilon}=\frac{i}{2l}\epsilon ,\ 
\gamma^a D_a \tilde{\epsilon} = -\frac{3}{4l^2} \epsilon .
\end{\eq}

Correspondingly, the reduced $\mathcal{N}=2$ CS action $S_{\rm CS}^{\rm r}$ and SYM action $S^r_{YM}$
are invariant under the transformation:
\begin{eqnarray}
\delta X_i     &=& -\frac{i}{2} ( \bar{\epsilon}\gamma_i \lambda -\bar{\lambda} \gamma_i \epsilon ) , \NN \\ 
\delta \sigma  &=&  \frac{1}{2} ( \bar{\epsilon}\lambda -\bar{\lambda} \epsilon ) , \NN \\ 
\delta \lambda &=& \frac{1}{2} \gamma^{ij}\epsilon f_{ij} -D\epsilon +\gamma^i \epsilon [ X_i ,\sigma ] +\frac{1}{l}\sigma \epsilon ,
                    \NN \\
\delta \bar{\lambda} &=&  \frac{1}{2} \gamma^{ij}\bar{\epsilon}f_{ij} +D\bar{\epsilon},
                         -\gamma^i \bar{\epsilon} [ X_i , \sigma ] +\frac{1}{2l}\sigma \bar{\epsilon} ,
                    \NN \\
\delta D       &=& -\frac{1}{2} \bar{\epsilon} \gamma^i [ X_i ,\lambda ]
                   -\frac{1}{2} [ X_i , \bar{\lambda} ] \gamma^i \epsilon
                   +\frac{i}{2} [ \bar{\epsilon}\lambda +\bar{\lambda}\epsilon ,\sigma ]
                   +\frac{1}{l}( \bar{\epsilon}\lambda +\bar{\lambda}\epsilon ) ,
\end{eqnarray}
where $f_{ij}$ is the reduced version of the field strength
\begin{eqnarray}
f_{ij}
&=& \frac{1}{2}\epsilon_{ijk} \Biggl[ \frac{2}{l}X_k  -\frac{i}{2}\epsilon_{klm} [\ X_l ,X_m\ ] \Biggr].    
\end{eqnarray}
Note that this supersymmetry is off-shell 
and therefore the localization works well.
Moreover, the reduced $\mathcal{N}=2$ SYM action is written in the following $Q$-exact form 
\begin{eqnarray}
 \bar{\epsilon}\epsilon S_{YM}^r
&=& \delta_{\bar{\epsilon}} \delta_\epsilon \Tr \Bigl[ \frac{1}{2}\bar{\lambda}\lambda -2D\sigma \Bigr]. 
\end{eqnarray}

\subsection{Matter sector}
Next we consider the supersymmetric transformation of the chiral multiplet $(\phi, \psi, F)$  on $S^3$.
The transformation is constructed in \cite{Kapustin:2009kz} for the canonical $R$-charge assignment and extended to general $R$-charge in
\cite{Jafferis:2010un, Hama:2010av}.
We suppose that the matter chiral multiplet is in the bi-fundamental representation 
under the $U(N_1) \times U(N_2)$ gauge group.  
The Lagrangian is
\begin{eqnarray}\label{chiralb}
\mathcal{L}_{kin} 
&=&  \Tr\Biggl[
     D_i \bar{\phi} D^i \phi +\bar{\phi}( \sigma_A -\sigma_B )^2 \phi 
    +\frac{i(2q-1)}{l} \bar{\phi}(\sigma_A -\sigma_B ) \phi \nonumber \\
&&  +\frac{q(2-q)}{l^2}\bar{\phi}\phi  +i\bar{\phi}(D_A -D_B ) \phi  +\bar{F}F \nonumber \\
&&  -i\bar{\psi}\gamma^i D_i \psi +i\bar{\psi}(\sigma_A -\sigma_B )  \psi -\frac{2q-1}{2l}\bar{\psi}\psi 
     +i\bar{\psi}(\lambda_A -\lambda_B ) \phi -i\bar{\phi}( \bar{\lambda}_A -\bar{\lambda}_B )\psi  
    \Biggr]. \nonumber \\
\end{eqnarray}
The action is invariant under the supersymmetric transformation: 
\begin{eqnarray}
\delta \phi       &=& \bar{\epsilon}\psi,  \NN \\
\delta \bar{\phi} &=& \epsilon \bar{\psi},  \NN \\
\delta \psi       &=& i\gamma^a \epsilon D_a \phi +i\epsilon\sigma_{AB}\phi +2qi\tilde{\epsilon}\phi +\bar{\epsilon}F, \NN \\
\delta \bar{\psi} &=& i\gamma^a \bar{\epsilon} D_a \bar{\phi} +i\bar{\phi}\sigma_{AB}\bar{\epsilon}
                       +2qi\bar{\phi}\tilde{\bar{\epsilon}} +\bar{F}\epsilon, \NN \\
\delta F          &=& \epsilon (i\gamma^a D_a \psi -i\sigma_{AB}\psi -i\lambda\phi ) -i(2q-1)\tilde{\epsilon}\psi, \NN \\
\delta \bar{F}    &=& \bar{\epsilon} (i\gamma^a D_a \bar{\psi} -i\bar{\psi}\sigma_{AB} -i\bar{\phi}\bar{\lambda} ) 
                      -i(2q-1)\tilde{\bar{\epsilon}}\bar{\psi},
\end{eqnarray}
where $\sigma_{AB}\equiv \sigma_A -\sigma_B$ and $q$ is $R$-charge of scalar field $\phi$

The action of the reduced model for the matter is
\begin{eqnarray}
S_{kin}^r 
&=& \Tr\Biggl[
     ( X_i \phi -\phi Y_i )^\dag  ( X_i {\phi} -\phi Y_i )
     +\bar{\phi}( \sigma_A -\sigma_B )^2 \phi 
    +\frac{i(2q-1)}{l} \bar{\phi}(\sigma_A -\sigma_B ) \phi \nonumber \\ 
&&    +\frac{q(2-q)}{l^2}\bar{\phi}\phi   +i\bar{\phi}(D_A -D_B ) \phi  +\bar{F}F \nonumber \\
&&  -\bar{\psi}\gamma^i (X_i \psi -\psi Y_i ) +i\bar{\psi}(\sigma_A -\sigma_B )  \psi -\frac{q-2}{l}\bar{\psi}\psi 
     +i\bar{\psi}(\lambda_A -\lambda_B ) \phi -i\bar{\phi}( \bar{\lambda}_A -\bar{\lambda}_B )\psi 
    \Biggr]. \nonumber \\
\end{eqnarray}
This is invariant under the following supersymmetric transformation
\begin{eqnarray}
\delta \phi       &=& \bar{\epsilon}\psi,  \NN \\
\delta \bar{\phi} &=& \epsilon \bar{\psi},  \NN \\
\delta \psi       &=& \gamma^i \epsilon D^r_i \phi +i\epsilon \sigma_{AB}\phi 
                      -\frac{q}{l}\epsilon\phi +\bar{\epsilon}F, \NN \\
\delta \bar{\psi} &=& \gamma^i \bar{\epsilon} D^r_i \bar{\phi} +i\bar{\phi}\sigma_{AB}\bar{\epsilon}
                       -\frac{q}{l} \bar{\phi}\bar{\epsilon} +\bar{F}\epsilon, \NN \\
\delta F          &=& \epsilon \left( \gamma^i D^r_i \psi  -i\sigma_{AB}\psi -i\lambda\phi \right) 
                       +\frac{q-2}{l} \epsilon \psi, \NN \\
\delta \bar{F}    &=& \bar{\epsilon} \left( \gamma^i D^r_i \bar{\psi} -i\bar{\psi}\sigma_{AB} -i\bar{\phi}\bar{\lambda} \right) 
                      +\frac{q-2}{l} \bar{\epsilon}\bar{\psi}. 
\end{eqnarray}
where we define $D_i^r \phi := X_i \phi -\phi Y_i   $.

The reduced matter action is also written in $Q$-exact form as
\begin{eqnarray}
\bar{\epsilon}\epsilon S_{kin}^r
&=& \delta_{\bar{\epsilon}} \delta_\epsilon 
    \Tr \Bigl[ \bar{\psi}\psi -2i\bar{\phi}(\sigma_{AB} )\phi +\frac{2(q-1)}{l}\bar{\phi}\phi \Bigr]. 
\end{eqnarray}

\section{Gauge fixing}\label{gaugefixing}
There are zero-eigenvalue in the matrix (\ref{kmatrix}) associated to gauge modes. 
In this appendix, we consider gauge-fixing of these modes.
The BRST transformation  is defined by
\begin{eqnarray}
\delta \left( -\frac{2}{l}L_i +X_i  \right) &=& -i \Bigl[\ -\frac{2}{l}L_i +X_i , c \Bigr], \\
\delta \left( \bar{\sigma} +\phi        \right) &=& -i \Bigl[\ \bar{\sigma} +\phi       , c \Bigr], \\
\delta c                                    &=& i \{c, c\}, \\
\delta b                                    &=& B, \\
\delta B                                    &=& 0.
\end{eqnarray}
The gauge-fixing action plus the ghost action is
\bea
S_{GF+FP} = \delta \Tr(bG)= \Tr\left( BG +b\delta G \right).
\eea

Here $G$ is the gauge-fixing function  
\begin{eqnarray}
G &=& [\ \sigma_0 ,\phi\ ] +\frac{2}{l} [\ L_i , X_i\ ], 
\end{eqnarray}
which is same as the eigenmodes associated to the zero eigenvalue.
Integration over the $B$ field generates the delta function constraint $\delta(G)$  and actually fix the gauge modes.
However, the integration measure of the gauge mode is normalized as $\int d G/\sqrt{G}$,
the normalization  the delta-function $\delta(G)=\frac{1}{\sqrt{G}}\delta(G/\sqrt{G})$ gives additional factor: 
\bea\label{normalization}
\prod_{s,t} \prod_{\alpha\in\Delta} \prod_{J=|j_s -j_t|}^{j_s +j_t} \prod_{m=-J}^{J}
    \left\{ \left( \frac{2}{l}\right)^2 J(J+1) +(\alpha\cdot\sigma )^2    \right\}^{-1/2}
=\frac{1}{\sqrt{\det{\Delta_{\sigma ,X^0}^2} } }.
\eea

Next we evaluate the Fadeev-Popov determinant. 
The BRST transformation of $G$ is
\bea  
\delta G
&=& -i[\ \sigma_0 , [\ \sigma_0 ,c\ ]\ ] +i \left( \frac{2}{l}\right)^2 [\ L_i ,[\ L_i ,c\ ]\ ]
\end{eqnarray}
and  the ghost action can be written as 
\begin{eqnarray}
\Tr\left( b\delta G \right)
&=& i\Tr\left( -b [\ \sigma_0 , [\ \sigma_0 ,c\ ]\ ] +\left( \frac{2}{l}\right)^2 b[\ L_i ,[\ L_i ,c\ ]\ ] \right) \\
&=& i\Tr\left( [\ \sigma_0 ,b\ ] [\ \sigma_0 ,c\ ] +\left( \frac{2}{l}\right)^2 b[\ L_i ,[\ L_i ,c\ ]\ ] \right).
\end{eqnarray}
The ghost fields are also expanded in terms of the scalar fuzzy sphere harmonics:
\begin{eqnarray}
&&c=\sum_{s,t}c^{(s,t)} =\sum_{s,t}    \sum_{J=|j_s -j_t|}^{j_s +j_t} \sum_{m=-J}^J 
                   \hat{Y}_{Jm(j_s j_t)} \otimes  c_{Jm}^{(s,t)} ,  \nonumber \\           
&&b=\sum_{s,t} b^{(s,t)} =    \sum_{J=|j_s -j_t|}^{j_s +j_t} \sum_{m=-J}^J 
                   \hat{Y}_{Jm(j_s j_t)} \otimes  b_{Jm}^{(s,t)} .           
\end{eqnarray}
Then we obtain the ghost action:
\begin{eqnarray}
\Tr\left( b\delta G \right)
&=& i\sum_{s,t} \Tr\left(  [\ \sigma_0 ,b\ ]^{(s,t)} [\ \sigma_0 ,c\ ]^{(t,s)} 
                         +\left( \frac{2}{l}\right)^2 b^{(s,t)} L_i \circ L_i \circ c^{(t,s)} \right) \nonumber \\
&=& i\sum_{s,t}\sum_{\alpha ,J,m} 
      \Biggl[  (\alpha\cdot\sigma )^2 b_{Jm}^{\alpha (s,t)\dag} c_{Jm}^{\alpha (s,t)} 
                         +\left( \frac{2}{l}\right)^2 J(J+1) b_{Jm}^{\alpha (s,t)\dag} c_{Jm}^{\alpha(s,t)} \Biggr] \NN \\
\end{eqnarray}
Integrating out $b_{Jm}^{\alpha (s,t)}$ and $c_{Jm}^{\alpha(s,t)}$, the 1-loop determinant for the ghosts is
\bea\label{fpdet}
\det{\Delta_{ghosts}}
&=& \prod_{s,t} \prod_{\alpha\in\Delta} \prod_{J=|j_s -j_t|}^{j_s +j_t} \prod_{m=-J}^{J}
    \left\{ \left( \frac{2}{l}\right)^2 J(J+1) +(\alpha\cdot\sigma )^2    \right\} \NN \\
&=& \det{\Delta_{\sigma ,X^0}^2} .
\eea
We can see that the one-loop determinant of $(\sigma, X^0)$, (\ref{normalization}) and (\ref{fpdet}) 
are canceled out with each other.

\section{Detailed calculation of  the 1-loop determinants}
\label{sec:det}
In this appendix, we present the detail explanation  for  degeneracy counting of the one-loop determinants
  and derivation of   (\ref{rymdeter}) and (\ref{mattldet}).

\subsection{Gauge sector}
First of all, we estimate the one-loop determinant of the reduced YM action.
The bosonic one-loop determinant consists of two parts.  
\begin{itemize}
\item $X_{Jm\rho}^{(s,t)}|_{\rho =+1}$
\begin{eqnarray}
\det{\Delta_X^2}|_{\rho =+1}
&=& \prod_{s,t} \prod_{\alpha\in\Delta} \prod_{J=|j_s -j_t|}^{j_s +j_t} \prod_{m=-(J+1)}^{J+1}
    \left\{ \left( \frac{2}{l}\right)^2 (J+1)^2 +(\alpha\cdot\sigma )^2    \right\} \nonumber \\
&=& \prod_{\alpha\in\Delta_+}  \prod_{s,t}\prod_{J=|s-t|/2}^{j_s +j_t} 
    \left\{ \left( \frac{2}{l}\right)^2 (J+1)^2 +(\alpha\cdot\sigma )^2    \right\}^{2(2J+3)} .\nonumber 
\end{eqnarray}
If we take the limit $\frac{n}{\nu}\ \RA\ \infty$, the determinant becomes
\begin{eqnarray}
\det{\Delta_X^2}|_{\rho =+1}
&=& \prod_{\alpha\in\Delta_+}  \prod_{s,t}\prod_{J=|s-t|/2}^{\infty} 
    \left\{ \left( \frac{2}{l}\right)^2 (J+1)^2 +(\alpha\cdot\sigma )^2    \right\}^{2(2J+3)} . 
\end{eqnarray}
In order to simplify the products, we count the degeneracy number of $(s,t)$ giving a same value of $J$.
First, note that possible values of $(s-t)$ giving a same $J$ is $-2J , -2J+2 ,\cdots ,2J$.
Next, the number of $(s,t)$ giving a fixed value of $|s-t|=m$ is $(\nu -m)$.
Thus, we obtain the degeneracy number as
\begin{\eqa}
\sum_{m=-2J,m:\rm even}^{2J} (\nu -m) = (2J+1)\nu -2J(J+1) \quad &{\rm for}& \ 2J={\rm even} \nonumber \\ 
\sum_{m=-2J,m:\rm odd}^{2J} (\nu -m)  = (2J+1)\nu -2J(J+1) \quad &{\rm for}& \ 2J={\rm odd} . \NN \\
\end{\eqa}
Therefore, a part of the bosonic one-loop determinant is rewritten as
\begin{eqnarray}
\det{\Delta_X^2}|_{\rho =+1}
&=& \prod_{\alpha\in\Delta_+} \prod_{J=0,\ 2J+1\in \mathbf{Z}}^{\infty} 
    \left\{ \left( \frac{2}{l}\right)^2 (J+1)^2 +(\alpha\cdot\sigma )^2    \right\}^{2(2J+3)\{ (2J+1)\nu -2J(J+1) \}} 
     \nonumber \\
&=& \prod_{\alpha\in\Delta_+}  \prod_{n=1}^{\infty} 
    \left\{ \left( \frac{n+1}{l}\right)^2  +(\alpha\cdot\sigma )^2    \right\}^{2(n+2)\{ n\nu -(n-1)(n+1)/2 \}} \nonumber \\
&=& \prod_{\alpha\in\Delta_+}  \prod_{n=1}^{\infty} 
    \left\{ \left( \frac{n+1}{l}\right)^2  +(\alpha\cdot\sigma )^2    \right\}^{2n(n+2)\nu -(n+2)(n-1)(n+1) } \nonumber \\
&=& \prod_{\alpha\in\Delta_+}  \left\{ \left( \frac{1}{l}\right)^2  +(\alpha\cdot\sigma )^2    \right\}^{-2}
    \prod_{n=1}^{\infty} 
    \left\{ \left( \frac{n}{l}\right)^2  +(\alpha\cdot\sigma )^2    \right\}^{2(n+1)(n-1)\nu -n(n+1)(n-2)}.
\label{X_plus}
\end{eqnarray}
The other part can be evaluated in a similar way.
\item $X_{Jm\rho}^{(s,t)}|_{\rho =-1}$\\
\begin{eqnarray}
\det{\Delta_X^2}|_{\rho =-1}
&=& \prod_{s,t} \prod_{\alpha\in\Delta} \prod_{J=|j_s -j_t|-1}^{j_s +j_t-1} \prod_{m=-J}^{J}
    \left\{ \left( \frac{2}{l}\right)^2 (J+1)^2 +(\alpha\cdot\sigma )^2    \right\} \nonumber \\
&=& \prod_{\alpha\in\Delta_+} 
    \prod_{n=1}^{\infty} 
    \left\{ \left( \frac{n-1}{l}\right)^2  +(\alpha\cdot\sigma )^2    \right\}^{2n(n-2) \nu -(n-1)(n-2)(n+1)} \nonumber \\
&=& \prod_{\alpha\in\Delta_+} \left\{ (\alpha\cdot\sigma )^2 \right\}^{-4\nu}
    \prod_{n=1}^{\infty} 
    \left\{ \left( \frac{n}{l}\right)^2  +(\alpha\cdot\sigma )^2    \right\}^{2(n+1)(n-1) \nu -n(n-1)(n+2)} . 
\label{X_minus}
\end{eqnarray}

Next we count the degeneracy of the ferminonic one-loop determinant.
\item $\lambda |_{\kappa =+1}$
\begin{eqnarray}
\det{\Delta_\lambda |_{\kappa =+1}}
&=& \prod_{s,t} \prod_{\alpha\in\Delta} \prod_{J=|j_s -j_t|}^{j_s +j_t} \prod_{m=-(J-1/2)}^{J+1/2}
    \left\{ \frac{2}{l}\left( J+1 \right) -i(\alpha\cdot\sigma )     \right\} \nonumber \\
&=& \prod_{\alpha\in\Delta_+} \prod_{n=1}^\infty 
   \left\{ \left( \frac{n+1}{l} \right)^2 +(\alpha\cdot\sigma )^2     \right\}^{n\nu (n+1) -(n-1)(n+1)^2 /2} \nonumber \\
&=& \prod_{\alpha\in\Delta_+}  \left\{ \left( \frac{1}{l} \right)^2 +(\alpha\cdot\sigma )^2  \right\}^{-1}
    \prod_{n=1}^\infty 
   \left\{ \left( \frac{n}{l} \right)^2 +(\alpha\cdot\sigma )^2     \right\}^{\nu n(n-1) -n^2 (n-2) /2} .
\label{lambda_plus}
\end{eqnarray}

\item $\lambda |_{\kappa =-1}$
\begin{eqnarray}
\det{\Delta_\lambda |_{\kappa =-1}}
&=& \prod_{s,t} \prod_{\alpha\in\Delta} \prod_{J=|j_s -j_t|-1/2}^{j_s +j_t-1/2} \prod_{m=-J}^{J}
    \left\{ \frac{2}{l}\left( -J-\frac{1}{2} \right) -i(\alpha\cdot\sigma )     \right\} \nonumber \\
&=& \prod_{s,t} \prod_{\alpha\in\Delta_+} \prod_{J=|j_s -j_t|-1/2}^{j_s +j_t-1/2} 
   \left\{ \left( \frac{2}{l}\right)^2 \left( J+\frac{1}{2} \right)^2 +(\alpha\cdot\sigma )^2   \right\}^{2J+1} \nonumber \\
&=& \prod_{\alpha\in\Delta_+} \prod_{n=1}^\infty 
   \left\{ \left( \frac{n-1}{l}\right)^2 +(\alpha\cdot\sigma )^2     \right\}^{n\nu (n -1) -(n-1)^2 (n+1)/2} \nonumber \\
&=& \prod_{\alpha\in\Delta_+} 
   \prod_{n=1}^\infty 
   \left\{ \left( \frac{n}{l}\right)^2 +(\alpha\cdot\sigma )^2     \right\}^{n(n +1)\nu -n^2 (n+2)/2} .
\label{lambda_minus}
\end{eqnarray}

\end{itemize}

Combining these result (\ref{X_plus}), (\ref{X_minus}), (\ref{lambda_plus}) and (\ref{lambda_minus})
we obtain in the $\nu\ \RA\ \infty$ limit:
\begin{\eqa}
\frac{ \det \Delta_{\lambda} |_{\kappa =+1} \cdot \det \Delta_{\lambda} |_{\kappa =-1} } 
     { \sqrt{\det \Delta^2_{ X}|_{\rho =+1}} \cdot \sqrt{\det \Delta^2_{ X}|_{\rho =-1}}   }
&=& \prod_{\alpha\in\Delta_+} \left\{ (\alpha\cdot\sigma )^2 \right\}^{2\nu}
    \prod_{n=1}^\infty  \left\{ \left( \frac{n}{l}\right)^2 +(\alpha\cdot\sigma )^2     \right\}^{2\nu}\NN \\ 
&=& \prod_{\alpha\in\Delta_+} \left\{ (\alpha\cdot\sigma )^2 \right\}^{2\nu}
    \prod_{n=1}^\infty \left( \frac{n}{l}\right)^{4\nu} 
   \left\{ \frac{ n^2 +l^2 (\alpha\cdot\sigma )^2 }{n^2}    \right\}^{2\nu}\NN \\ 
&=& \prod_{\alpha\in\Delta_+} \left\{ (\alpha\cdot\sigma )^2 \right\}^{2\nu}
     (2\pi l)^{2\nu} \left\{ \frac{\sinh{(\pi l(\alpha\cdot\sigma ))}}{\pi l\alpha\cdot\sigma}  \right\}^{2\nu} \NN \\
&=& \prod_{\alpha\in\Delta_+} 
      \left( 2\sinh{(\pi l(\alpha\cdot\sigma ))} \right)^{2\nu} .
\end{\eqa}
From the second line to the third line in the above equations, we used the following formula:
\begin{\eqa}
&& \prod_{n=1}^\infty \left( \frac{n^2 +x^2}{n^2} \right) = \frac{\sinh{(\pi x)}}{\pi x}, \\
&& \prod_{n=1}^\infty n^2 =e^{2\zeta^\prime (0)}          = 2\pi ,\\
&& \prod_{n=1}^\infty c = e^{\zeta (0) \log{c}} =\frac{1}{\sqrt{c}} .
\end{\eqa}

\subsection{Matter sector}
Next, we study the matter sector. 
The degeneracy counting for the matter one-loop determinant is parallel to the vector multiplet.
The calculation for bosonic part is following: 
\begin{itemize}
\item $(\phi ,\bar{\phi})$\\
\begin{eqnarray}
&& \det{\Delta_{\phi}^2} \nonumber \\
&=& \prod_{s,t} \prod_{I_1, I_2} \prod_{J=|j_s -j_t|}^{j_s +j_t} \prod_{m=-J}^{J}
    \Biggl[ \left( \frac{2}{l}\right)^2 J(J+1) +(\sigma_{I_1} -\tilde{\sigma}_{I_2})^2
           +\frac{i(2q-2)}{l}(\sigma_{I_1} -\tilde{\sigma}_{I_2}) +\frac{q(2-q)}{l^2}
    \Biggr] \nonumber \\
&=& \prod_{s,t} \prod_{I_1, I_2} \prod_{J=|j_s -j_t|}^{j_s +j_t} \prod_{m=-J}^{J}
    \Biggl[ \left( \frac{2}{l}\right)^2 J(J+1) 
           +\left( \sigma_{I_1} -\tilde{\sigma}_{I_2} +i\frac{q}{l} \right)
            \left( \sigma_{I_1} -\tilde{\sigma}_{I_2} +i\frac{(q-2)}{l} \right)
    \Biggr] \nonumber \\
&=& \prod_{I_1, I_2} \prod_{n=1}^{\infty} 
    \Biggl[ \left( \frac{n+1}{l} -\frac{q}{l} +i(\sigma_{I_1} -\tilde{\sigma}_{I_2})  \right)
            \left( \frac{n-1}{l} +\frac{q}{l} -i(\sigma_{I_1} -\tilde{\sigma}_{I_2})  \right)
    \Biggr]^{n^2 \nu -n(n-1)(n+1)/2}. 
\end{eqnarray}
The contributions from the fermionic part are given by 
\item $\left. (\psi ,\bar{\psi}) \right|_{\kappa =+1}$\\
\begin{eqnarray}
\det{\Delta_\psi |_{\kappa =+1}}
&=& \prod_{s,t} \prod_{I_1 ,I_2} \prod_{J=|j_s -j_t|}^{j_s +j_t} \prod_{m=-(J-1/2)}^{J+1/2}
    \left\{ \frac{2}{l} J +i(\sigma_{I_1} -\tilde{\sigma}_{I_2} ) -\frac{q-2}{l}   \right\} \nonumber \\
&=&  \prod_{I_1 ,I_2}  \prod_{n=1}^\infty
    \left\{ \frac{n+1}{l} +i(\sigma_{I_1} -\tilde{\sigma}_{I_2} ) -\frac{q}{l}   \right\}^{n\nu (n+1) -(n-1)(n+1)^2 /2}. 
\end{eqnarray}

\item $\left. (\psi ,\bar{\psi}) \right|_{\kappa =-1}$\\
\begin{eqnarray}
\det{\Delta_\psi |_{\kappa =-1}}
&=& \prod_{s,t} \prod_{\alpha\in\Delta} \prod_{J=|j_s -j_t|-1/2}^{j_s +j_t-1/2} \prod_{m=-J}^{J}
    \left\{ \frac{2}{l}\left( -J-\frac{3}{2} \right) +i(\sigma_{I_1} -\tilde{\sigma}_{I_2} ) -\frac{q-2}{l}     \right\} \nonumber \\
&=& \prod_{I_1 ,I_2} \prod_{n=1}^{\infty} 
    \left\{ -\frac{n-1}{l} +i(\sigma_{I_1} -\tilde{\sigma}_{I_2} ) -\frac{q}{l}     \right\}^{n\nu (n -1) -(n-1)^2 (n+1)/2} .
\end{eqnarray}

\end{itemize}

Thus, in the $\nu\rightarrow\infty$, we obtain
\begin{eqnarray}
\frac{\det{\Delta_\psi |_{\kappa =+1}} \det{\Delta_\psi |_{\kappa =-1}} }{\det{\Delta_{\phi}^2}}
&=& \prod_{I_1 ,I_2}  \prod_{n=1}^\infty \Biggl[
     \frac{\frac{n+1}{l} -\frac{q}{l} +i(\sigma_{I_1} -\tilde{\sigma}_{I_2}) }
          {\frac{n-1}{l} +\frac{q}{l} -i(\sigma_{I_1} -\tilde{\sigma}_{I_2}) }
      \Biggr]^{n\nu} \nonumber \\
&=& \prod_{I_1 ,I_2} s_b^\nu (i-iq -l(\sigma_{I_1} -\tilde{\sigma}_{I_2} ) ).
\end{eqnarray}

\bibliographystyle{JHEP}

\bibliography{EKref,localization,SUSYlattice}
\end{document}